\documentclass[12pt]{elsarticle}

\usepackage{url}
\usepackage{hyperref}
\usepackage[anythingbreaks]{breakurl}
\usepackage{rotating}

\usepackage{mathrsfs,mathtools,latexsym,eucal}
\usepackage{amscd,amsfonts,amssymb,amsmath}
\usepackage{tikz-cd}
\usepackage{tkz-graph}
\usetikzlibrary{arrows,shapes}

\usepackage{graphicx,graphics,color}
\usepackage{geometry,setspace}
\usepackage{enumerate}
\usepackage{booktabs}
\usepackage{scalerel}
\usepackage[utf8]{inputenc}
\usepackage[active]{srcltx}

\usepackage{float} 

\usepackage{listings}

\definecolor{dkgreen}{rgb}{0,0.6,0}
\definecolor{gray}{rgb}{0.5,0.5,0.5}
\definecolor{mauve}{rgb}{0.58,0,0.82}

\begin{document}

\let\today\relax
\makeatletter
\def\ps@pprintTitle{%
    \let\@oddhead\@empty
    \let\@evenhead\@empty
    \def\@oddfoot{\footnotesize\itshape
         { }}%
    \let\@evenfoot\@oddfoot
    }
\makeatother

\hyphenation{Fi-gu-re}\hyphenation{ele-ments}


 \title{A Fixed-Mass multifractal approach \\ for unweighted complex networks} 

 \author[mymainaddress]{Pablo Pav\'on-Dom\'inguez$^{1,}$\corref{mycorrespondingauthor}\fnref{myfootnote}\fnref{myfootnote2}}
 \ead{pablo.pavon@uca.es. }
 \cortext[mycorrespondingauthor]{Corresponding author}
 \fntext[myfootnote]{Escuela Superior de Ingenier\'ia, Avenida de la Universidad de C\'adiz, 11519 Puerto Real, C\'adiz, Spain.}
 \fntext[myfootnote2]{Both authors contributed equally to the work.}

 \author[mysecondaryaddress]{Soledad Moreno-Pulido$^{1,}$\fnref{myfootnote}\fnref{myfootnote2}}
 \ead{soledad.moreno@uca.es}

 \address[mymainaddress]{Dept. of Mechanical Engineering and Industrial Design, University of C\'adiz, Spain.}
 \address[mysecondaryaddress]{Dept. of Mathematics, University of C\'adiz, Spain.}

 \begin{abstract}

Complex networks have been studied in recent years due to their relevance in biological, social and technical real systems, such as the world wide web, social networks and biochemical interactions. One of the most current features of complex networks is the presence of (multi-)fractal properties. In spite of the amount of contributions that have been developed in fractal and multifractal analyisis, all of them have focused on the adaptation of Fixed-size algorithms (FSA) to complex networks, mostly box-counting and sandbox procedures. In this manuscript, a Fixed-mass algorithm (FMA) is adapted to explore the (multi-)fractality of unweighted complex networks is first proposed on the basis on a box-counting procedure. Thus, the aim of this work is to explore the applicability of a FMA to study fractal and multifractal characteristics of unweighted complex networks. After describing the proposed FMA for complex networks, it is tested in several kinds of complex networks such as deterministic, synthetic and real ones. Results suggest that FMA is able to adequately reproduce the fractal and multifractal properties of synthetic complex networks (scale-free, small-world and random). In addition, the algorithm is used for the mutifractal description of real complex networks. The main advantage of FMA over FSA is that the minimum configuration of boxes (ideal) for completely covering the network is known. Therefore, results are quite precise when a reasonable covering configuration is obtained, allowing the reconstruction of the Mass exponent, $\tau$, and the Dimension function, $D(q)$, with proper goodness-fits.


 \end{abstract}

 \maketitle

\section{Introduction}

Complex networks are of enormous relevance for summarizing the complexity of many social, biological and technological phenomena \cite{Barabasi02} (the world wide web, social networks and biochemical interactions). Some of them exhibit fractal features such as self-organization, self-similarity, attractor, small-world phenomenon and scale-free degree (see \cite{Song2005, Gallos2012, Watanabe2015, Bogdan2016, Bogdan2018}). From the innovating work of Song et al. \cite{Song2005}, several studies dealing with the verification of fractality and the determination of the fractal dimension of complex networks have been developed (e.g. \cite{Kawasaki10, Song2006}). In fact, self-similarity has become the third basic characteristic of complex networks, along with the small-world character and scale-free property \cite{Liu17}.

Complex networks can be excellently synthesized by mathematical models \cite{Wen2018}, they being usually represented as graphs. Graphs are comprised by two basic elements: nodes and edges. Whereas nodes are the elements of the system, edges represent the interactions among nodes. A complex network can be embedded in the Euclidean space such as real world networks (e.g. transport networks). However, in abstract networks, geography may not be relevant since their embedded space is not generally Euclidean \cite{Kim2007}. The absence of a geometric support allows to emphasize the topological relationships among nodes \cite{Liu2015}, but hinders the traditional box-counting procedure \cite{feder88}, which is a basic step for fractal analysis. Generally, a box-counting procedure consists in covering the object with non-overlapping pieces which will be known as boxes.

In complex networks, the concept of box-counting deals with the search of the minimum number of boxes to cover the network to ensure a non-overlapping covering. Generally, the way of minimally covering a complex network with equivalent boxes is a NP (non-deterministic polynomial-time) hard problem \cite{Furuya11}, that is, the computational time to solve the problem could be very long and non-defined. The first approximation to the estimation of the fractal dimension in complex networks was developed by Song et al. \cite{Song2005}. In that work, authors extended the concept of ``box'' for a certain group of nodes in a complex network. A modified version of this method was proposed by Kim et al. \cite{Kim2007}, providing an easier implementation of the algorithm. Later, Song et al. \cite{Song2007} proposed a greedy colouring algorithm for box covering procedures, as well as burning algorithms, which are generally simpler and computationally faster. In a burning algorithm, a random node acts as a seed of a box and its unburned neighbour nodes are assigned to this box. Many subsequent works have proposed modifications in the box-counting algorithm e.g.\cite{Kim2007b,Rozenfeld2007,Zhou2007,Gao2008,Sun&Zhao2014}.

The subsequent step was the study of complex networks from multifractal techniques. For their study, some algorithms have been adapted to analyse the multifractality of networks. Furuya and Yakubo \cite{Furuya11} first evidenced that a single fractal dimension could be not enough for completely characterize the multifractal behaviour in scale-free networks. These authors demonstrated analytical and numerically by means of applying a CBB algorithm that multifractal features are due to large fluctuations of local node density. As same as these authors, Wang et al. \cite{Wang2012} studied the multifractal properties of several families of both synthetic complex networks (scale-free, small world and random) and real networks (protein-protein interactions). They obtained the dimension function through averaging a box-counting procedure over a large number of times. This implemented procedure was called  box-covering multifractal algorithm for complex networks, which was improved by Li et al. \cite{Li14}. The improvement proposed was to find an optimal covering configuration of the network with a minimal amount of boxes, which is a more reasonable extension of the traditional multifractal analysis. An alternative for covering the network is the so-called ‘sand-box’ procedure \cite{Tel1989,VICSEK1990,VicsekFamily1990}. Based on the aforementioned procedure, Liu et al., \cite{Liu2015} developed an extension of the sand-box algorithm  for characterizing multifractal properties in complex networks.

Summarizing, the generalization of different algorithms for the fractal and multifractal analysis of complex networks has been very prolix in the last decade. The improvements of the algorithms have been based on both their adaptation to the characteristics of the different types of complex networks and the proposal of box-counting algorithms as effective as possible \cite{Wei2013, Song2015, Wei2016, Xue17, Liu17}. However, all these contributions are based on a Fixed-size approach (FSA), but no work has been conducted on the use of a Fixed-mass approach (FMA) for complex networks. In the FMA, the box covering is performed with a box of a certain mass $N_{box}$, which consider the $k$th nearest neighbours and compute the smallest radius $R_{min}$ to include them in it, as $N_{box}$ increases \cite{Theiler90}. FMA were first introduced by \cite{Termonia83} and followed by \cite{Badii1984, Badii1985} for the determination of the fractal dimension and the generalized Renyi dimensions of complex dynamic systems. Later, the FMA was also employed for the evaluation of the entire dimension spectrum $f(\alpha)$ by Badii and Broggi \cite{BadiiBroggi88}.

As same as in FSA, NP hard also appears in FMA. This causing difficulty for obtaining a minimum configuration of boxes for completely covering the complex network without overlapping. But unlike the first, relevant information about the goodness of the optimal covering is obtained when a FMA is used in complex networks. That is, the theoretical minimum amount of boxes needed to completely cover the network is known. In addition, FMA is not affected by the main drawbacks of the FSA, avoiding the quasi-empty boxes problem \cite{Grassberger1988}. FMA provides accurate estimations of the multifractal parameters and exhibits some advantages that have already been highlighted in the analysis of different phenomena \cite{Hirabayashi1992,Mach95, Pastor97,DeBartolo00,Enescu2005,Behnia12}.

Therefore, the objective of this work is to explore and evaluate the applicability of the FM approach for characterizing fractal and multifractal features in unweighted complex networks. For this purpose, an adaptation of the box-counting FM algorithm to unweighted complex networks is introduced. Later, the proposed algorithm is applied to deterministic, synthetic and real complex networks. Results evince that FMA is suitable for studying fractal and multifractal properties in unweighted complex networks by means of precise reconstruction of the dimension function, $D(q)$.

\section{Methods}

\subsection{Multifractal analysis}

The multifractal characterization of a measure can be obtained by the following function $\Phi$ defined by \cite{Falconer94}:
\begin{equation}\label{Phi}
	\Phi(q,\tau)=\left\langle \sum_{i=1}^N p_i^{q-1} \varepsilon_i^{-\tau} \right\rangle,
\end{equation}
where $p_i$ and $\varepsilon_i$ are the probability of mass and the size of the object in the part $i$, respectively; $N$ is the amount of separated parts in the object and the brackets denote the average of the measure over different realizations \cite{Pastor97,Mach95}. 
The function $\Phi$ represents the pairs $q$ moment of the measure and the $\tau$ moment of the size. For any measure, it is assumed that this function collapses onto a single constant value for fine enough partitions:

\begin{equation}\label{mediaconstante}
	\left\langle \sum_{i=1}^N p_i^{q-1} \varepsilon_i^{-\tau} \right\rangle=k.
\end{equation}

This expression determines a dependence between the values of $\tau$ and $q$ whenever $p_i$ or $\varepsilon_i$ is fixed. On the one hand, when the size $\varepsilon_i=\varepsilon$ is fixed, the values of $\tau$ depends on the values of $q$ and equation \eqref{mediaconstante} yields to
\begin{equation}\label{fs}
	\left\langle \sum_{i=1}^{N(\varepsilon)} p_i^{q-1} \right\rangle \sim \varepsilon^{\tau},
\end{equation}
where $N(\varepsilon)$ is the number of parts with size $\varepsilon$ needed to cover the whole support in which the measure is defined.

On the other hand, when the mass $p_i=p$ is fixed, the values of $q$ depends on the values of $\tau$ and equation \eqref{mediaconstante} yields to
\begin{equation}\label{fm}
	\left\langle \sum_{i=1}^{N(p)} \varepsilon_i^{-\tau} \right\rangle \sim p^{1-q},
\end{equation}
where $N(p)$ is the number of parts of measure $p$.

According to this, there are two kinds of multifractal box-counting ways to estimate the parameters $\tau$ and $q$, the so-called Fixed-size (FS) \cite{Grassberger83} and the Fixed-mass (FM) approaches \cite{Termonia83,Badii1984,BadiiBroggi88}. In the FM approach, the parameter $q$ depends on the values of $\tau$ $\bigl(q=q(\tau)\bigr)$ and by simply taking logarithms in expression \eqref{fm}, it can be written as:

\begin{equation}\label{logfm}
	\ln\left(\left\langle \sum_{i=1}^{N(p)} \varepsilon_i^{-\tau} \right\rangle\right) \sim (1-q)\ln(p).
\end{equation}

And equation \eqref{logfm} is used to estimate the values of $(1-q)$ as the slope of the linear regressions of $\ln\left(\left\langle \sum_{i=1}^{N(p)} \varepsilon_i^{-\tau} \right\rangle\right)$ versus $\ln(p)$. The generalized dimension function $D(\tau)$, characterizes the multifractal properties and can be computed by equation \eqref{Dq}:
\begin{equation}\label{Dq}
	D(\tau)=\frac{\tau}{q(\tau)-1},\mbox{ for all } q\neq1.
\end{equation}

\subsection{Fixed mass algorithm for complex networks}

A FMA for (multi-)fractal analysis of complex networks is proposed in this section by means of a box-counting procedure. Hence, the main goal is to completely cover the network with the least amount of fixed-mass boxes, and then to compute the diameter of these boxes. In this work, this method is explored in unweighted networks.

\begin{figure}[H]
\centering
\includegraphics[width=\textwidth]{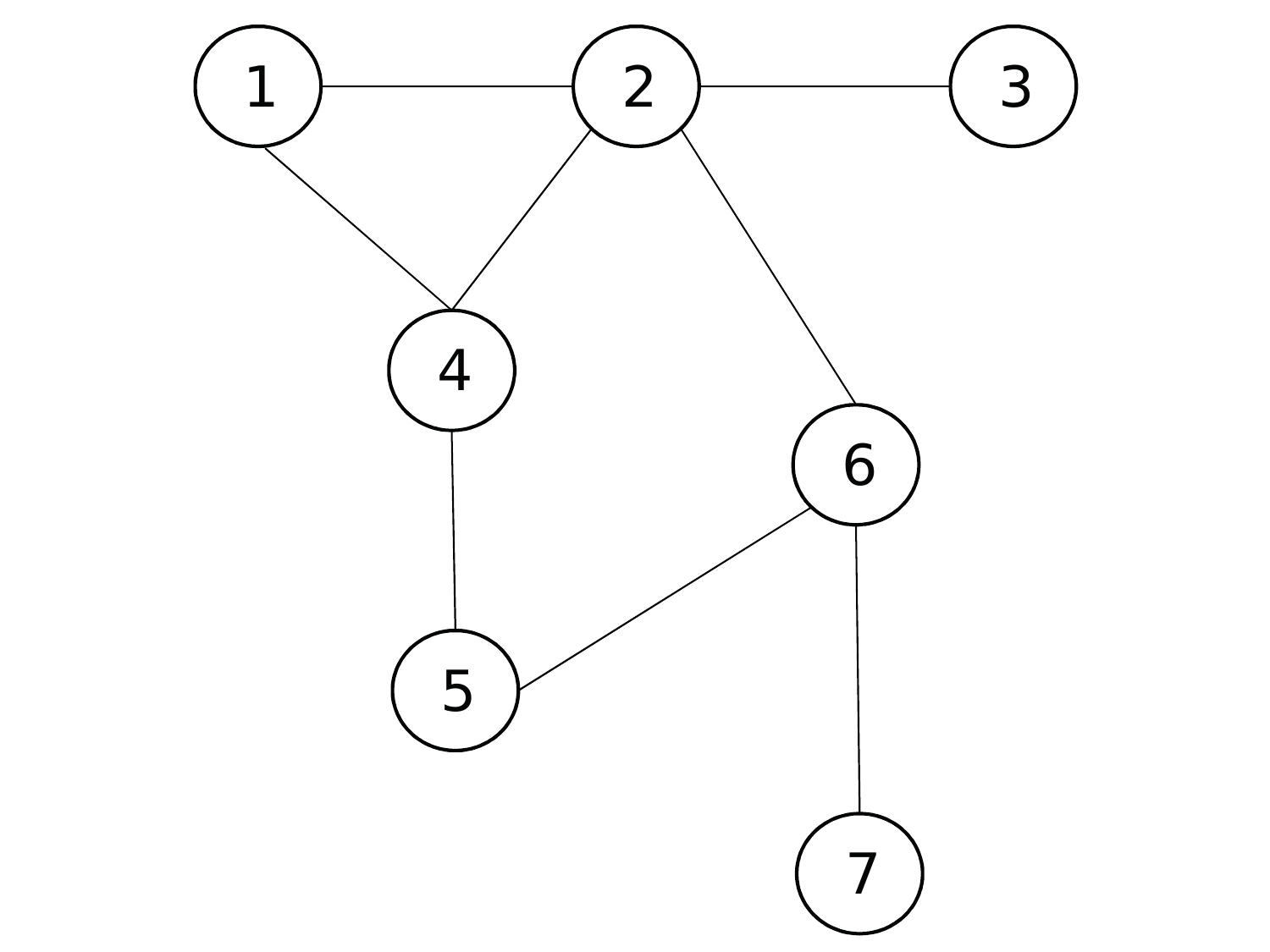}
\caption{Example Network.}
\label{figura_grafo}
\end{figure}

Unweighted networks show the connections between the nodes, but not consider the strength of the edges, i.e., the edges are not labeled with values. To represent complex networks with a total amount of nodes $N_{network}$, the \emph{adjacency matrix} $A_{N_{network}\times N_{network}}$ is usually computed. This is a square matrix such that $a_{ii}=0$; $a_{ij}=1$ if $i\neq j$ and the node $i$ is directly connected with the node $j$; and $a_{ij}=\infty$ otherwise. Throughout this work, only undirected networks will be considered, which implies that the adjacency matrix is a symmetric one. As an example, for the network in Figure \ref{figura_grafo}, the adjacency matrix is
$$A=\left(
       \begin{array}{ccccccc}
         0 &          1 &     \infty &          1 &     \infty &     \infty &     \infty \\

         1 &          0 &          1 &          1 &     \infty &          1 &     \infty \\

    \infty &          1 &          0 &     \infty &     \infty &     \infty &     \infty \\

         1 &          1 &     \infty &          0 &          1 &     \infty &     \infty \\

    \infty &     \infty &     \infty &          1 &          0 &          1 &     \infty \\

    \infty &          1 &     \infty &     \infty &          1 &          0 &          1 \\

    \infty &     \infty &     \infty &     \infty &     \infty &          1 &          0 \\

       \end{array}
    \right).$$

To compute the distance between all nodes that comprise an unweighted network, the \emph{shortest path matrix} $D_{N_{network}\times N_{network}}$ is considered, where $d_{ij}$ is the minimum number of edges connecting the nodes $i$ and $j$. As well as the adjacency matrix, the shortest path matrix is symmetric if the network is undirected. Throughout this work, only networks with one connected component will be considered. In this example, the shortest path matrix is
$$D=\left(
       \begin{array}{ccccccc}
         0 &          1 &          2 &          1 &          2 &          2 &          3 \\

         1 &          0 &          1 &          1 &          2 &          1 &          2 \\

         2 &          1 &          0 &          2 &          3 &          2 &          3 \\

         1 &          1 &          2 &          0 &          1 &          2 &          3 \\

         2 &          2 &          3 &          1 &          0 &          1 &          2 \\

         2 &          1 &          2 &          2 &          1 &          0 &          1 \\

         3 &          2 &          3 &          3 &          2 &          1 &          0 \\

       \end{array}
    \right).$$
The shortest matrix will be used as an input data for the multifractal FMA proposed. The FMA is introduced as follows:

\begin{enumerate}[(1)]

    \item Establish a wide enough set of values of $\tau$.

    \item Rearrange all the nodes that comprise the network in $t=1,\dots,T$ random sequences. For each sequence, nodes will be selected as a center of a box. According to the size of the networks, $T=200$ is considered in this work.

    \item Set the measure of the box, $N_{box}$, in the range $[2,N_{network}]$.

    \textbf{Remark 1:} The measure considered is the amount of nodes ($N_{box}$) inside each box instead of the probability mass $p_{box}=\displaystyle\frac{N_{box}}{N_{network}}$. Notice that the normalization with a constant factor $N_{network}$ does not change the results \cite{FmaDeBartolo06}.

    \item For each center of a box, find a subgraph comprised by $N_{box}-1$ nearest neighbours through the shortest path matrix $D$, and cover the nodes belonging to this box that have not been covered yet.

    \textbf{Remark 2:} The box considered contains exactly the fixed mass $N_{box}$ since the center also belongs to the box.

    \item  If every node in a box has been previously covered, then discard this empty box.

    \item Repeat steps $(4)$ and $(5)$ until all nodes of the network are covered. The number of boxes in this sequence will be denoted by $B(N_{box},t)$.

    \item Repeat steps $(4)$ to $(6)$ for all random sequences and retain the one with the minimum number of boxes $B_{min}$.

    \item For the minimal covering configuration, consider the diameter $\varepsilon_i$ of each box and calculate the partition sum $\displaystyle M(\varepsilon_i,\tau) = \frac{\displaystyle\sum_{i=1}^{B_{min}} \varepsilon_i^{-\tau}}{B_{min}}$.

    \textbf{Remark 3:} Notice that the diameter of a box is always lower than $N_{box}-1$.

    \item Finally, for different measures, $N_{box}$, repeat the steps $(4)$ to $(8)$, and then use $\ln(M\bigl(\varepsilon_i,\tau)\bigr)$ versus $\ln(N_{box})$ for calculating the linear regressions for each value of $\tau$. The slopes of these linear regressions are $(1-q)$.

    \textbf{Remark 4:} The proper values of $q$ should be selected in an appropriate range \cite{meneveau_sreenivasan_1991}. In general, $N_{network}$ is related to $10^q$. Throughout this paper and for comparison purposes, the range of $q$ will be $[-3,3]$.

\end{enumerate}

\begin{figure}[H]
\centering
\includegraphics[width=\textwidth]{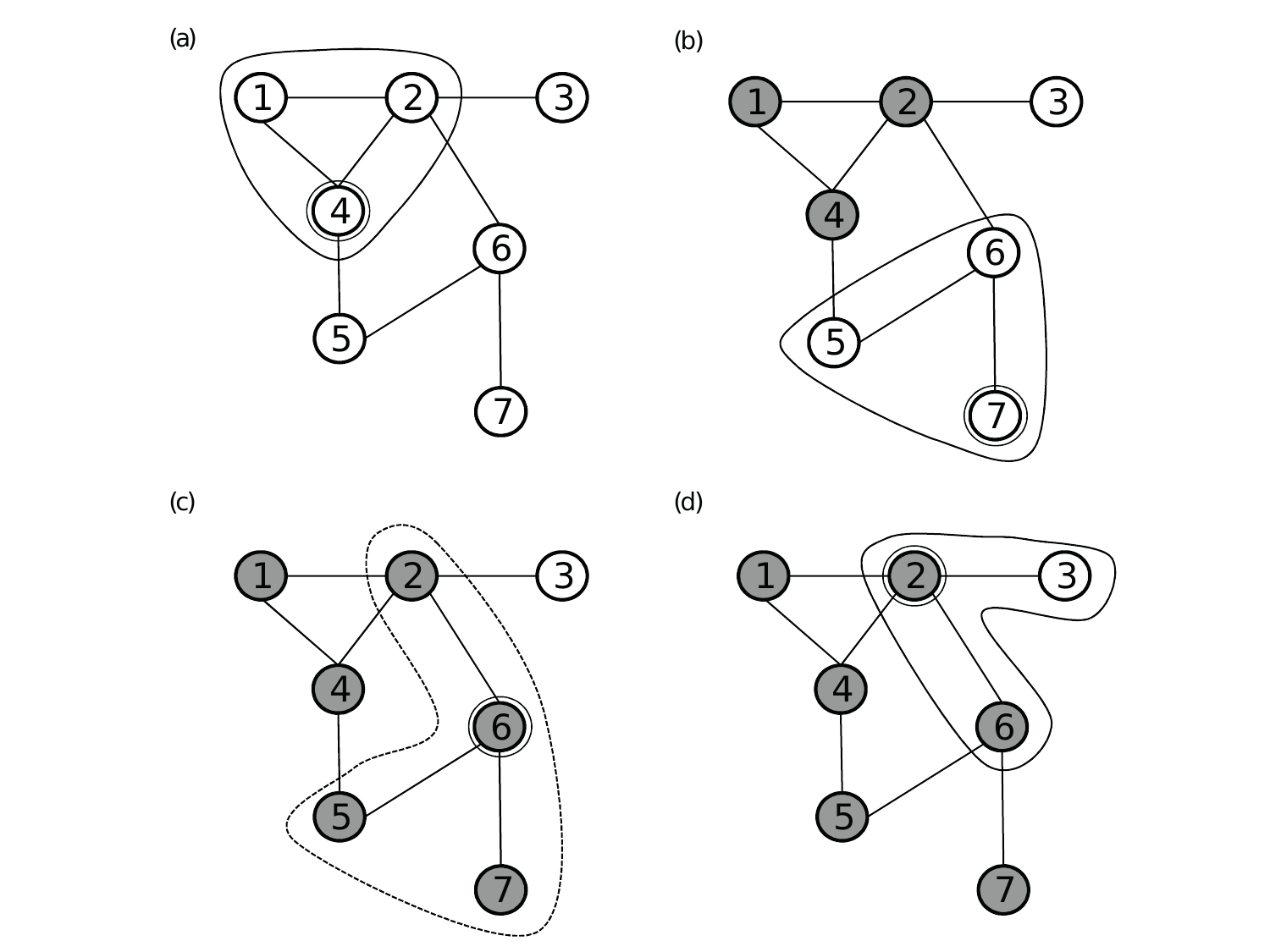}
\caption{\textbf{FMA Steps $\mathbf{(2)-(5)}$ for the example network for $\boldsymbol{N_{box}=3}$ and a rearrangement of the nodes given by $\mathbf{[4,\; 7,\; 6,\; 2,\; 1,\; 5,\; 3]}$:} (a) For the center $4$, the algorithm looks for a subgraph comprised by $2$ nearest neighbours. Since there are $3$ nearest neighbours $1,2,5$, the FMA randomly decided to select nodes $1$ and $2$, and then the first box is formed by $[1,2,4]$ with a diameter $1$. (b) For the center $7$, the algorithm looks for a subgraph comprised by $2$ nearest neighbours. Since the nearest neighbour is node $6$, the FMA picks it up and then searches for the nearest neighbours of node $6$, which are $2$ and $5$. Since node $5$ is not covered yet, the algorithm prioritises node $5$ over node $2$, and then the second box is formed by $[5,6,7]$ with a diameter $2$. (c) For the center $6$, the algorithm looks for a subgraph comprised by $2$ nearest neighbours. Since node $6$ and its neighbours $2,5$ and $7$ are already covered, the box is discarded. (d) For the center $2$, the algorithm looks for a subgraph comprised by $2$ nearest neighbours. Since the nearest neighbours are $1,3,4,6$ and node $3$ is not covered yet, the algorithm picks it up and then completes the box with another random neighbour already covered. In this case, the third box is comprised by $[2,3,6]$ with a diameter $2$. This FMA realization finishes with $3$ boxes to completely cover the network.}
\label{figura_FMA}
\end{figure}

A MATLAB code of the FMA is included as electronic supplementary material. See Figure \ref{figura_FMA} for an example of the FMA.

\section{Results}

\subsection{Deterministic networks}

The proposed FMA will be performed in deterministic networks to compare the results with the known analytic exponent mass function and fractal dimension. The goodness of this fitting will be tested through the following measure, which is inspired in the coefficient of determination in a polynomial regression fitting:

\begin{equation}\label{bondad_ajuste_FMA}
    G_f=1-\frac{S_r}{S_t},
\end{equation}
where
$$
    S_r=\displaystyle\frac{1}{n_{\tau}}\sum_{\tau}(q(\tau) - q_{\scaleto{FMA}{4pt}}(\tau))^2
$$
and
$$
    S_t=\displaystyle\frac{1}{n_{\tau}}\sum_{\tau}(q(\tau) - \overline{q})^2,
$$
the $n_{\tau}$ is the number of values of $\tau$ used in the fit; $q(\tau)$ and $q_{\scaleto{FMA}{4pt}}(\tau)$ represent the analytical and numerical results obtaining by the FMA, respectively. Finally, $\overline{q}$ is the average of $q(\tau)$. The nearer $G_f$ is to $1$, the better the fitting.

\subsubsection{Regular Ring}

A regular ring is a circular network comprised by $N_{network}$ and $N_{network}-1$ edges such that every node is only connected to its two neighbour nodes. In this work, a regular ring with $N_{network}=2500$ has been considered in order to apply the FMA proposed.

\begin{figure}[H]
 	\begin{center}
 		\includegraphics[width=\textwidth]{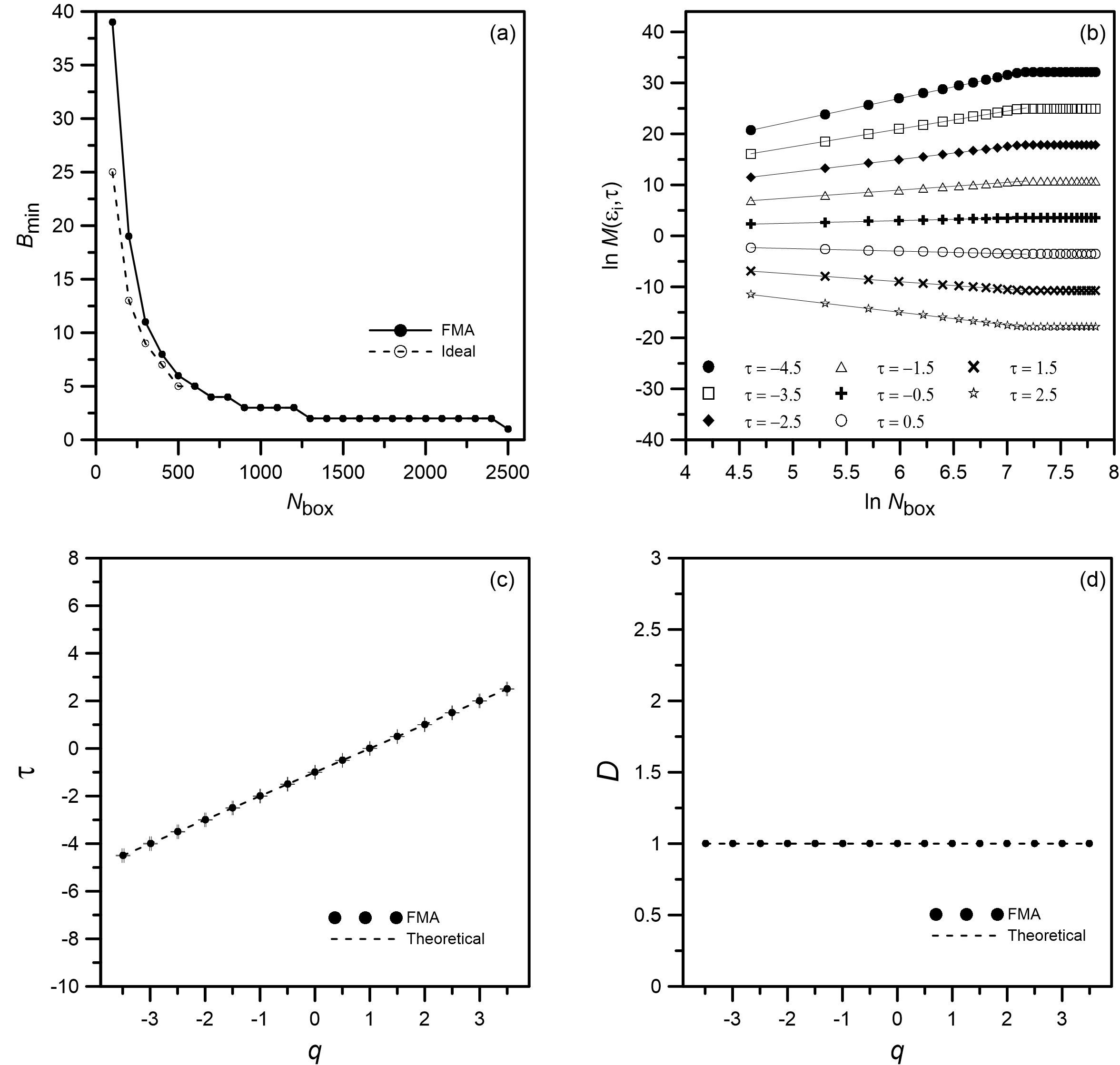}
         \caption{\textbf{Regular ring:} (a) Box-covering configuration. $B_{min}$ represents the minimum amount of boxes with a fixed $N_{box}$. (b) Partition functions. $\ln(M\bigl(\varepsilon_i,\tau)\bigr)$ versus $\ln(N_{box})$ for every value of $\tau$. (c) Exponent mass function.  (d) Generalised dimension function. }
         \label{Fig_01}
     \end{center}
 \end{figure}

After applying this algorithm to the regular ring, a minimal covering configuration is obtained. In Figure \ref{Fig_01} (a) (solid line) the minimum number of boxes versus the masses have been depicted. As expected, the shape of this plot follows a power law with a staircase effect for higher masses. Notice that for a mass $N_{box}=N_{network}$, the minimum number of boxes is $1$, for $\displaystyle\frac{N_{network}}{2}\leq N_{box} < N_{network}$, it is $2$, for $\displaystyle\frac{N_{network}}{3}\leq N_{box} < \frac{N_{network}}{2}$, it is $3$ and so on. In the FMA approach, the ideal amount of boxes can be known \emph{a priori} by the formula $N_{ideal\;box}=\displaystyle\frac{N_{network}}{N_{box}}$, as it is depicted in Figure \ref{Fig_01} (a) (dotted line). Hence, the closer the solid line is to the dotted line, the better will be the approximation. As seen in this Figure, the algorithm exhibits more divergence for lower masses than for the higher ones.

Figure \ref{Fig_01} (b) depicts the linear regressions of the partition functions for several values of $\tau$. It is evidenced that these functions are rather smooth and a clear linear range can be defined for linear fits. For higher masses, a strong flat region is found, which is related to the aforementioned saturation and staircase effect for masses higher than $\displaystyle\frac{N_{network}}{2}$. The linear fits performed to obtain the slopes $(1-q)$ of the regular ring have been determined in the range $N_{box} = [100, 1300]$. The coefficients of determination of theses linear fits are greater than $0.99$.

Multifractal results are shown in Figure \ref{Fig_01} (c) and (d). The plotting of the mass exponent function exhibits a linear behaviour. The error bars depicted are the errors to estimate the values of $q$, which are less than $0.03$. The theoretical $\tau = q-1$ function is depicted in a dotted line along with the FMA results. As seen, both functions almost overlap each other. Since $\tau$ is a linear function, the dimensions function is a constant equal to the slope of the exponent mass, which is 1 in this case. These results evince the capability of the proposed FMA to reproduce the non fractal nature of the regular ring network.

\subsubsection{Lattice graph}

In this lattice graph every node is connected to other $4$ neighbour nodes, except at the boundary, where nodes have less connections. In this work, a lattice with $N_{network}=2521$ and $N_{edges}=4900$ has been generated.

\begin{figure}[H]
 	\begin{center}
 		\includegraphics[width=\textwidth]{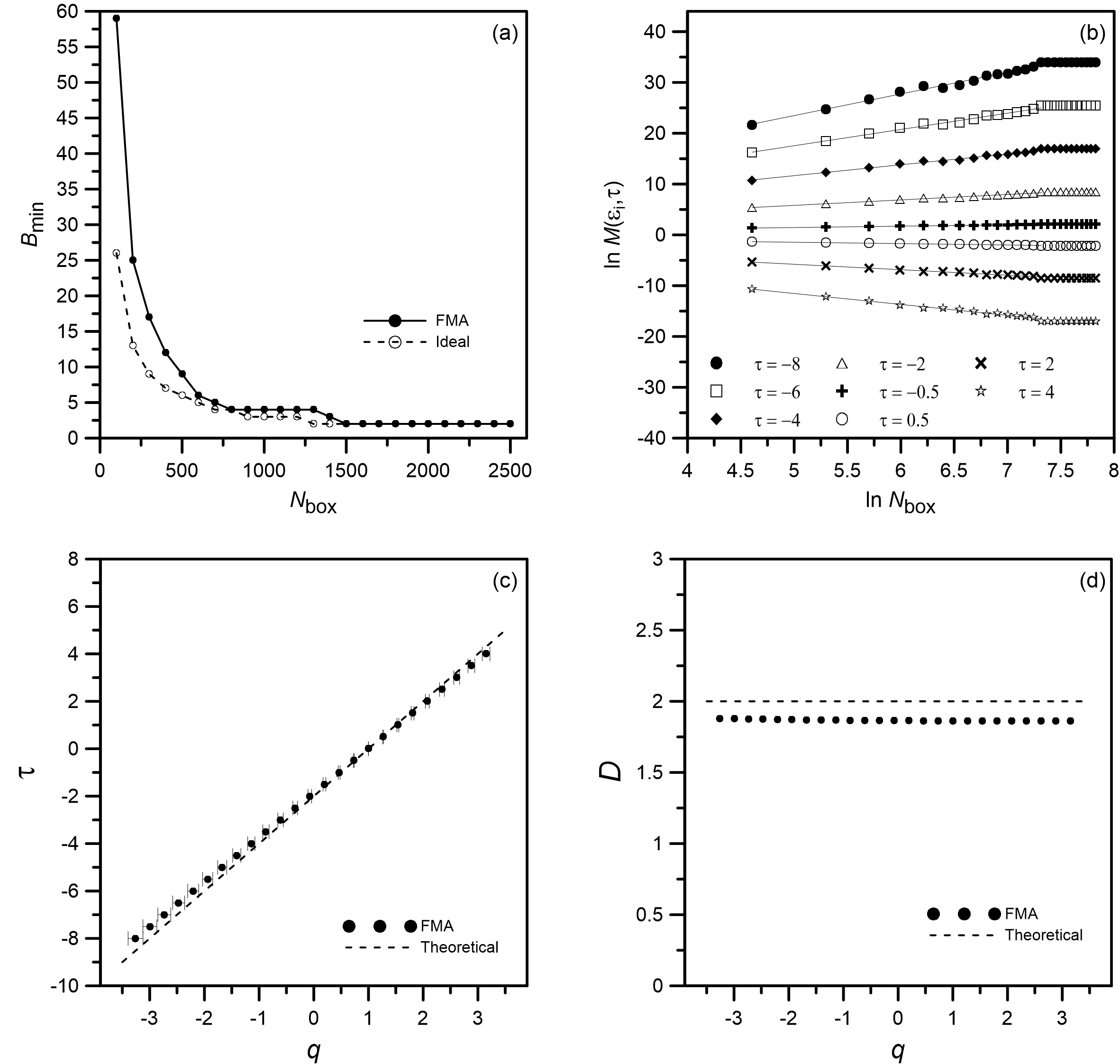}
         \caption{\textbf{Lattice graph:} (a) Box-covering configuration. $B_{min}$ represents the minimum amount of boxes with a fixed $N_{box}$. (b) Partition functions. $\ln(M\bigl(\varepsilon_i,\tau)\bigr)$ versus $\ln(N_{box})$ for every value of $\tau$. (c) Exponent mass function. (d) Generalised dimension function.}
         \label{Fig_02}
     \end{center}
 \end{figure}

The FMA has been performed to obtain the multifractal parameters of this network. Figure \ref{Fig_02} (a) exhibits a power law relationship between the masses and the minimum number of boxes. As in the regular ring, the ideal amount of boxes is depicted through a dotted line. For medium and higher masses, the FMA minimum configuration shows a staircase effect very similar to the ideal one.

Linear fits in the partition functions, with $R^2>0.98$, have been performed in the range $N_{box}=[100,1500]$, until the saturation process begins, as seen in Figure \ref{Fig_02} (b). On the other hand, the mass exponent function depicted in Figure \ref{Fig_02} (c) shows a linear behaviour. Error bars represent the estimation errors from the linear fits (lower than $0.28$). The theoretical $\tau = 2(q-1)$ function is depicted in a dotted line along with the FMA results. As seen, both functions approach each other, mainly for central values of $\tau$. Since $\tau$ is a linear function, the dimensions function is a constant equal to the slope of the exponent mass, which is 2 in this case. As seen in Figure \ref{Fig_02} (d), the results given by the FMA are a bit lower than the theoretical value $2$ due to the border effects and the finite size of the generated lattice graph.

According to the results observed in Figure \ref{Fig_01} and Figure \ref{Fig_02}, the capability of the FMA proposed has been shown to accurately characterize the non fractal behaviour of these deterministic complex networks.

\subsubsection{$(u,v)$-Flower}

Rozenfeld et al. \cite{Rozenfeld2007} developed a study about the $(u,v)$ flowers models. In the first generation of these deterministic models, a circular ring is established with a total amount of nodes and edges equal to $w = u + v$, where $1 < u\leq v$. The $n$th generation is obtained by removing the edges of the $(n-1)$th generation and by replacing them with two parallel paths with a length (number of edges) of $u$ and $v$ respectively. These authors also provided a formula for calculating both the degree exponent $\gamma = 1 + \displaystyle\frac{\ln(u+v)}{\ln(2)}$ and the fractal dimension $D_f = \displaystyle\frac{\ln(u+v)}{\ln(u)}$.

Later, Furuya and Yakubo \cite{Furuya11} explored the multifractality of complex networks through these $(u,v)$-flowers models. They proposed an analytic formula (Equation \eqref{tau_flower}) for computing the exponent mass of these networks and found that local node densities of the $(u,v)$ flower are distributed in a \emph{bifractal} way.
\begin{equation}\label{tau_flower}
\tau(q)= \left\{ \begin{array}{lcc}
             q\displaystyle\frac{\log(w/2)}{\log(u)} &   \mbox{if}  & q \geqslant \displaystyle \frac{\log(w)}{\log(2)} \\
             (q-1)\displaystyle\frac{\log(w)}{\log(u)} &  \mbox{if} & q < \displaystyle\frac{\log(w)}{\log(2)}
             \end{array}
   \right.
\end{equation}
In this work a $6$th generation of a $(2,2)$-flower has been analysed through the proposed FMA. This network has $N_{network} = 2732$ and $N_{edges} = 4096$.

\begin{figure}[H]
 	\begin{center}
 		\includegraphics[width=\textwidth]{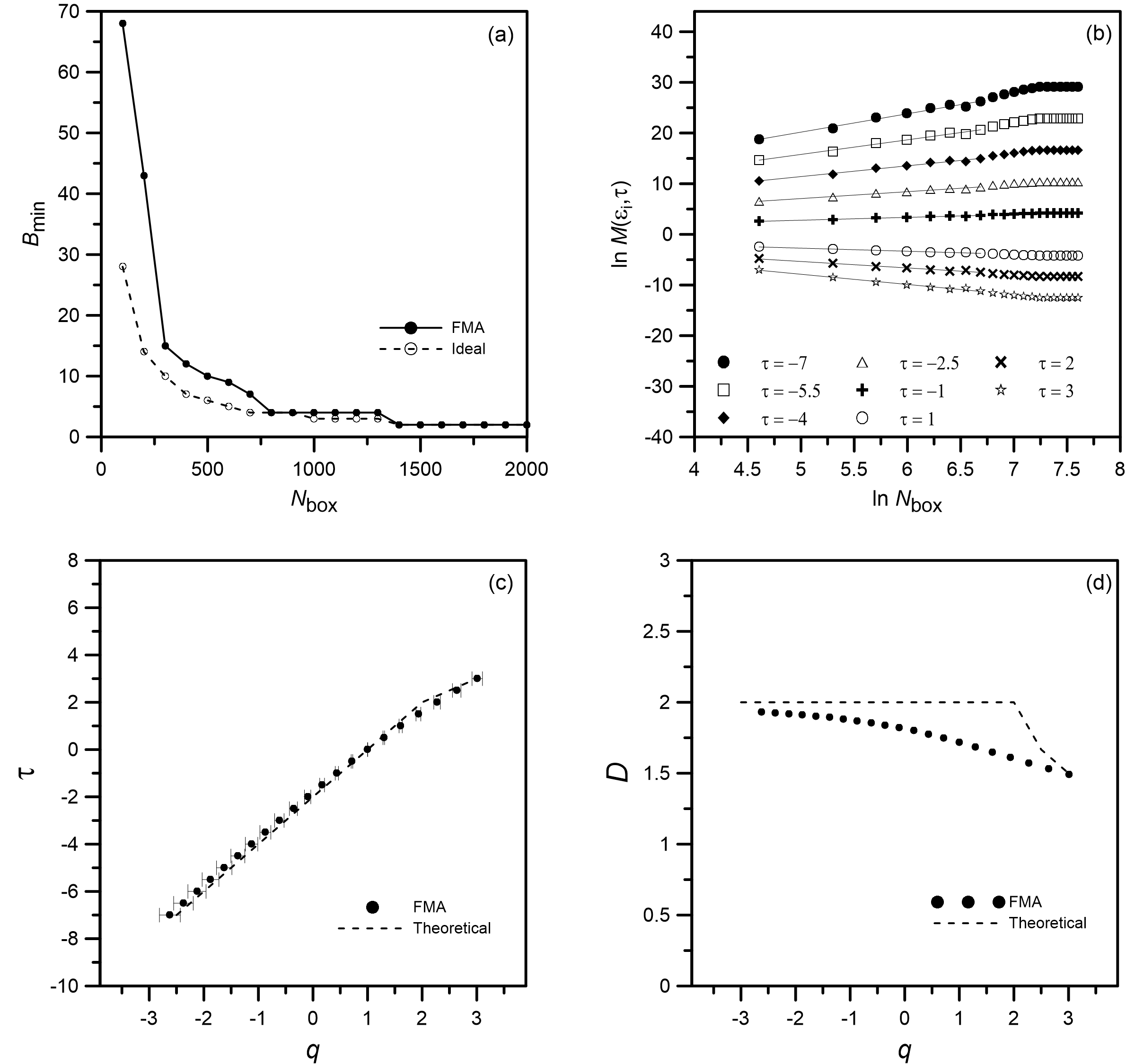}
         \caption{\textbf{(2-2)-flower:} (a) Box-covering configuration. $B_{min}$ represents the minimum amount of boxes with a fixed $N_{box}$. (b) Partition functions. $\ln(M\bigl(\varepsilon_i,\tau)\bigr)$ versus $\ln(N_{box})$ for every value of $\tau$. (c) Exponent mass function. (d) Generalised dimension function.}
         \label{Fig_03}
     \end{center}
 \end{figure}

Results depicted in Figure \ref{Fig_03} (a) show a power law distribution of the minimum number of boxes versus the masses $\bigl($see Figure \ref{Fig_03} (a)$\bigr)$. As before, the FMA tends to fit the ideal case represented in a dotted line. Figure \ref{Fig_03} (b) represents the partition functions in which the linear fit is performed in the range of masses $N_{box} = [100,800]$, these regressions show $R^2>0.98$. In Figure \ref{Fig_03} (c), the FMA estimation (bullets) of the mass exponent function is depicted along with the theoretical one (equation \eqref{tau_flower}, dotted line). The errors are represented through horizontal bars (lower than $0.19$) and the graphics almost overlap each other. Figure \ref{Fig_03} (d) depicts the generalized dimension function. Multifractal results provided by the FMA reproduces the characteristic bifractality of this network.

\begin{table}[H]
    \centering
    \begin{tabular}{lc}
    \toprule
         Networks                        & $G_f$ \\\midrule
         Linear ring                     & 0,999\\
         Lattice graph                   & 0,995\\
         $(u,v)$ flower                  & 0,994\\
         \bottomrule
    \end{tabular}
    \caption{Goodness of the fitting of FMA for deterministic networks.}
    \label{tabla_errores}
\end{table}

After studying the FMA proposed in deterministic networks, it is confirmed that this procedure reproduces with rather accuracy the non-fractallity of the regular ring and lattice and the bifractallity behaviour of the $(2,2)$-flower networks. Table \ref{tabla_errores} summarizes the goodness of the fitting proposed in equation \eqref{bondad_ajuste_FMA} for the three studied theoretical complex networks, accomplishing the criterion established by $G_f\approx1$.

\subsection{Synthetic networks}

\subsubsection{Scale-free networks}

The scale-free behaviour in complex networks has been studied in depth in fractal and multifractal framework \cite{Song2005, Song2006, Furuya11}. It is said that a network is scale-free when the  distribution of the node degrees follows a power law, ${\displaystyle P(k)\ \sim \ k^{\boldsymbol {-\gamma }}}$, where $P(k)$ is the fraction of nodes in the network having $k$ connections to other nodes, and the parameter $\gamma$ is the scaling exponent, which is accepted in the range $2 < {\displaystyle \gamma } < 3$ for scale-free networks.

In this work, the Barabasi-Albert model \cite{Barabasi99} is performed in order to generate scale-free networks. This model deals with two steps. First, we start with an initial `seed' identical for all the networks generated in this study. Nodes of the seed are required to form a connected graph. Second, a new node is added, which will be connected to node $i$ with a probability $p_i$ such that
$$
p_i = \displaystyle\frac{k_i}{\sum_j k_j},
$$
$k_i$ being the degree of node $i$. Due to the way these networks are generated, new nodes will be more likely to be connected to nodes with higher degree (hubs).

\begin{figure}[H]
 	\begin{center}
 		\includegraphics[width=\textwidth]{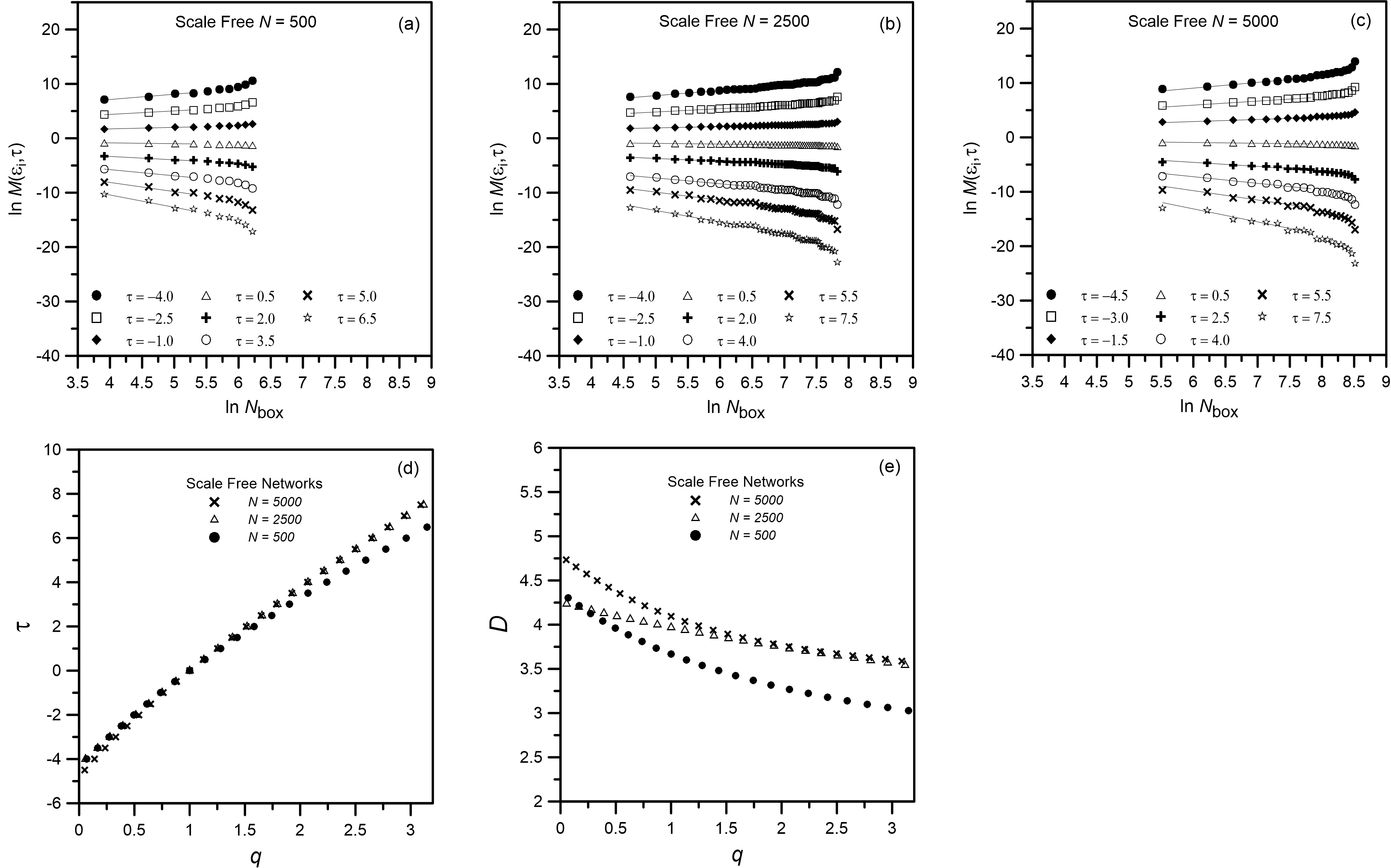}
         \caption{\textbf{Scale free networks:} Partition functions $\ln(M\bigl(\varepsilon_i,\tau)\bigr)$ versus $\ln(N_{box})$ for every value of $\tau$. (a) $N_{network}=500$. (b) $N_{network}=2500$. (c) $N_{network}=5000$. (d) Exponent mass functions. (e) Generalised dimension functions.}
         \label{Fig_04}
     \end{center}
 \end{figure}

The seed employed in this work is the same proposed by Wang et al. \cite{Wang2012} with adjacency matrix:
$$S=\left(
       \begin{array}{ccccc}
               0      & 1  & \infty      & \infty      & 1 \\
               1      & 0  & \infty      & 1      & \infty \\
               \infty      & \infty  & 0      & 1      & \infty \\
               \infty      & 1  & 1      & 0      & \infty \\
               1      & \infty  & \infty      & \infty      & 0 \\
       \end{array}
    \right).$$
Three scale-free networks have been considered in this work with $N_{network}$ equal to 500, 2500 and $5000$ nodes. The FMA is performed to compute the dimension function with appropriated values of $\tau$, to obtain the values of $q$ in the range $[-3,3]$. Figure \ref{Fig_04} (a-c) shows the partition functions of these networks. As seen, a linear region can be detected from smaller masses to $N_{network}/2$, in which linear fits have been performed with $R^2$ being higher than 0.97, 0.97 and 0.94, for scale-free networks of 500, 2500 and 5000 nodes, respectively. It should be noted that for masses higher than $N_{network}/2$, the partition functions do not follow a saturation behaviour as described for deterministic networks. Conversely, the relationship between ln$N_{box}$ versus ln$M$ is a potential-like behaviour which indicates that low increments in the mass of the box, implies large increases in its diameter. This fact may indicate that partitions functions obtained from the FMA proposed are capable of detecting the scale free behavior of complex networks.

In Figure \ref{Fig_04} (d) is evinced the multifractal behaviour of scale free networks, which is also observed from the decreasing dimension functions (Figure \ref{Fig_04} (e)). The multifractallity is explained for the different edge density in these networks due to the presence of hubs. Table \ref{tabla_sinteticas} summarizes the main parameters of these scale-free networks. These results are in agreement with the multifractallity detected in scale-free networks by means of FSA approaches \cite{Wang2012, Liu2015}.

\subsubsection{Small-world networks}

Complex networks exhibit a small-world behaviour when the distance between two randomly chosen nodes increases proportionally to the logarithm of the number of nodes. Actually, in these networks it is possible to link two separated nodes in a very few steps.

Watts and Strogatz \cite{Watts98} developed a model (WS model) for generating small-world networks by a rewiring process that starts with a regular network and finalizes with a random one. Thus, small-world networks retain both the high clustering coefficient of regular networks and the short average path length of the random ones. Later, Newman and Watts (1999) \cite{Newman99} improved and proposed a new model easier to compute (WS model). The NW model is followed here as same as in Liu et al. \cite{Liu2015} and it is equivalent to the WS model for small $p$ and sufficiently large $N_{network}$ \cite{Wang2012}, where $0 \leq p \leq 1$ and $N_{network} \gg K \gg \ln(N_{network}) \gg 1$. First, a ring regular with $N_{network}$ is generated in which each node is connected to its $K$ nearest neighbours. Secondly, a new edge between two non adjacent nodes can be added according to a probability $p$. In this work, the regular ring has been generated with $5000$ nodes by considering a value of $K=50$. The probabilities selected are $p=0,1\cdot10^{-5},2\cdot10^{-5}\mbox{ and }3\cdot10^{-5}$.

\begin{figure}[H]
 	\begin{center}
 		\includegraphics[width=\textwidth]{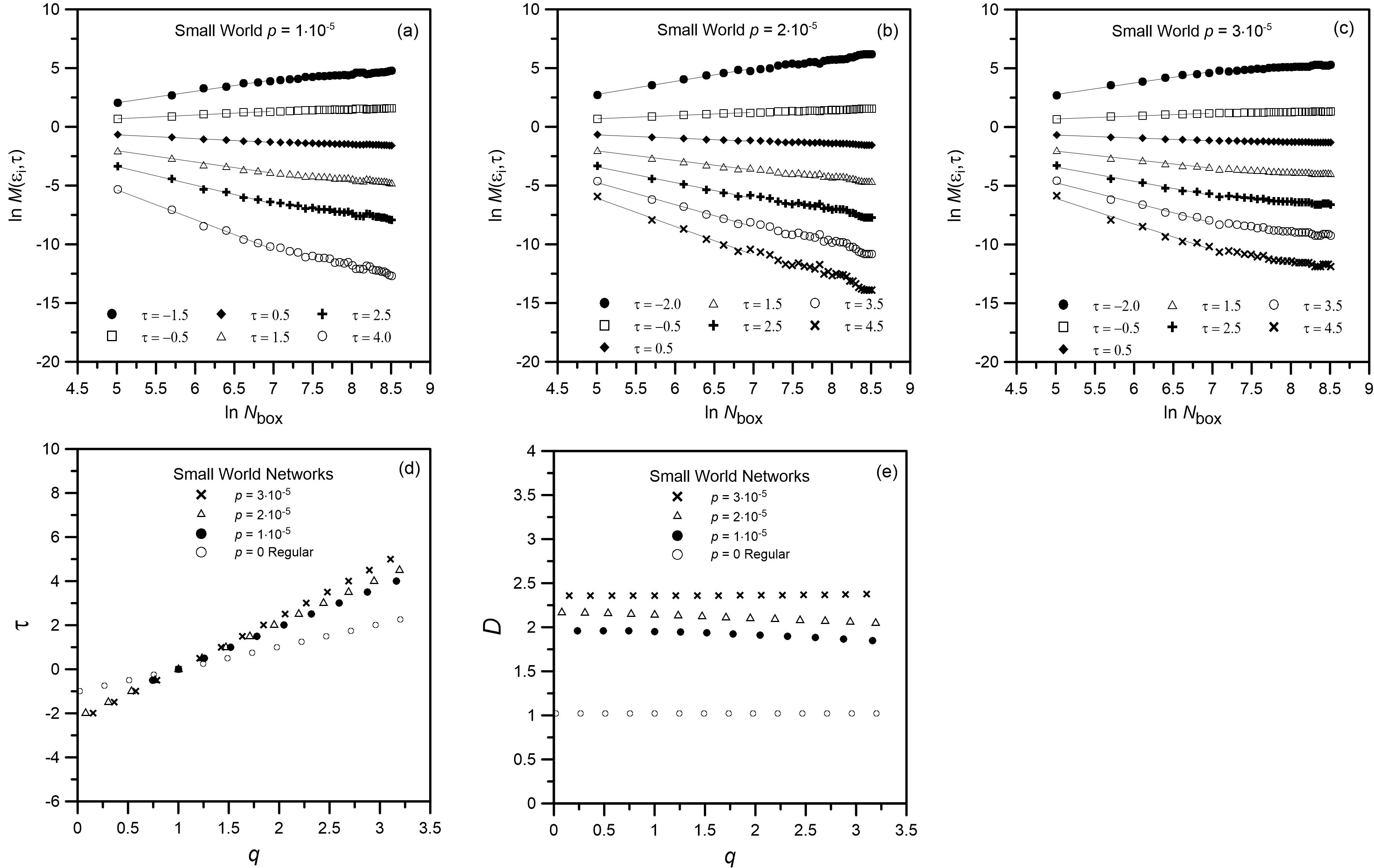}
         \caption{\textbf{Small World networks:} Partition functions $\ln(M\bigl(\varepsilon_i,\tau)\bigr)$ versus $\ln(N_{box})$ for every value of $\tau$.  (a) $p=1\cdot 10^{-5}$. (b) $p=2\cdot 10^{-5}$. (c) $p=3\cdot 10^{-5}$. (d) Exponent mass functions. (e) Generalised dimension functions.}
         \label{Fig_05}
     \end{center}
 \end{figure}

FMA is then applied to these complex networks and partition functions are calculated and depicted in Figure \ref{Fig_05} (a-c). The linear scaling behaviour is determined in the range $[450,1025]$ for the three studied small world complex networks. As seen, the maximum mass selected for linear fits is around $N_{network}/5$. The goodness of the linear fits is ensured with $R^2$ higher than 0.97, 0.91 and 0.94 for $p=1\cdot10^{-5},2\cdot10^{-5}\mbox{ and }3\cdot10^{-5}$, respectively.

Two relevant findings can be obtained from the partitions functions. On the one hand, the linear behaviour is found for a narrower range of masses and, on the other hand, the partition functions exhibit an increasing saturation process by following a staircase effect, which is more evident for positive values of $\tau$. The initial masses are discarded because the networks keep the features of the original network (regular ring with $K = 50$ neighbours). As the masses increase, the scale behavior of the small world begins to be detected, until a maximum mass value around  $N_{network}/5$ is reached, in which the partition functions show a staircase effect. This is caused by the low value of the diameter of these networks (see Table \ref{tabla_sinteticas}) and, therefore, the scarce amount of possible diameters of the boxes that exist in this type of networks. Thus, the partition functions obtained from the application of FMA allows to detect small world features in complex networks.

In Figure \ref{Fig_05}(d) and (e), the exponent mass and the generalized dimension functions of these small world networks are depicted along with the original ring selected to generate them. This ring exhibits the same non-fractal behaviour studied in the subsection of deterministic networks. Dimensions functions of small world networks practically are independent of $q$. In addition, according to Table \ref{tabla_sinteticas}, multifractallity in this networks is not obvious, but the higher the values of $p$, the greater the amount of edges are added, and thus, the value of $D(q=0)$ increases. Results obtained from FMA are in agreement with the non-clear multifractallity detected in \cite{Wang2012, Liu2015}.

\subsubsection{Random networks}

Finally, three random networks were generated by appliying the Erd\"{o}s-R\'{e}nyi (ER) random graph model \cite{Erdos60}. In this model, both a initial amount of isolated nodes and a probability $p$ are fixed. Then, it is evaluated if every pairs of nodes are connected by means of the probability of linkage $p$. Finally, a disconnected graph is obtained. Then, the major connected part of the graph is extracted and it is used for the further FMA.

\begin{figure}[H]
 	\begin{center}
 		\includegraphics[width=\textwidth]{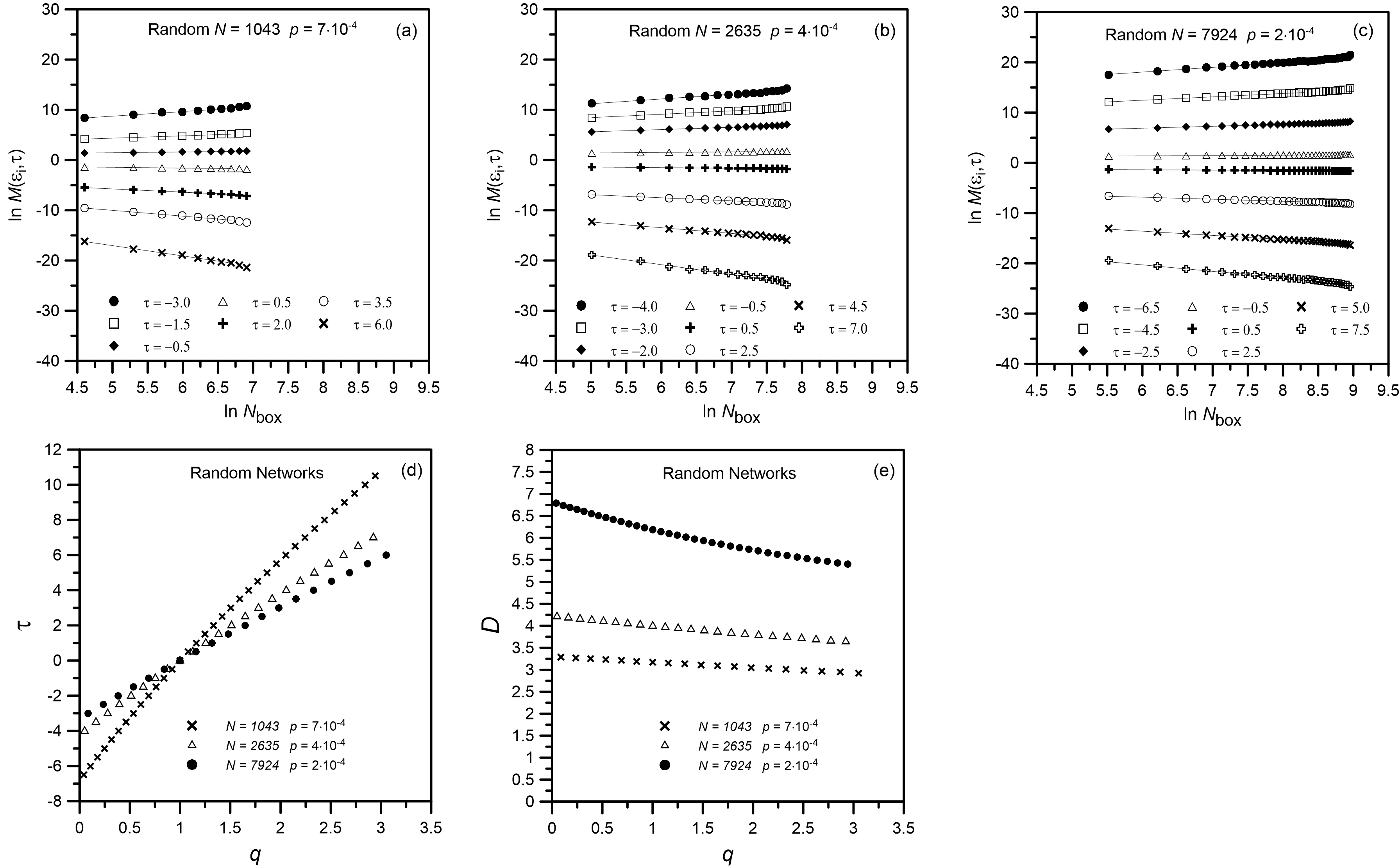}
         \caption{\textbf{Random networks:} Partition functions $\ln(M\bigl(\varepsilon_i,\tau)\bigr)$ versus $\ln(N_{box})$ for every value of $\tau$. (a) $N_{network}=1043$, $p=7\cdot 10^{-4}$. (b) $N_{network}=2635$, $p=4\cdot 10^{-4}$. (c) $N_{network}=7924$, $p=2\cdot 10^{-4}$. (d) Exponent mass functions. (e) Generalised dimension functions.}
         \label{Fig_06}
     \end{center}
 \end{figure}

Here, the three ER random graph models considered are comprised by $2000$ nodes, $4000$ nodes, and $20000$ nodes. The linkage probability is $p = 7\cdot 10^{-4}$, $p=4\cdot 10^{-4}$ and $p=2\cdot 10^{-4}$, respectively. This results in a largest connected component for these random graph with $N_{network} = 1043$, $2635$ and $7924$, respectively. These complex networks are analyzed with the proposed FMA and partition functions are depicted in Figure \ref{Fig_06}(a-c), respectively. As seen, the shape of these functions is different than the aforementioned scale-free and small-world partition functions. These functions seem to grow continuously and do not saturate, because of the randomness component in these networks, which produces a continuous increase in the radius of the boxes as the mass increases. The mass-radius relationship does not follow a specific behavior as in the case of complex networks affected with scale-free or small-world features.

Figure \ref{Fig_06} (d) and (e) represents the exponent mass and the dimension functions of these random networks. As seen, the greater the amount of nodes, the higher the fractal dimension and tends to multifractallity (see also Table \ref{tabla_sinteticas}), which is in agreement with results obtained by \cite{Wang2012}. According to these authors, the unexpected trend to multifractallity occurs because, during the generating process, nodes are randomly connected with probability $p$, and few hubs may exist. This is rather evident for random networks with great amount of nodes. In \cite{Liu2015}, the expected monofractallity of random networks is more evident because they used an average of the results over a set of 100 random networks.

\begin{table}[H]
\small
    \centering
    \begin{tabular}{lccccccc}
\toprule
 Network & $N_{network}$ & $N_{edges}$ &  Diameter &  $D(q=0)$ & $D(q=3)$ & $\Delta(D_q)$ & $\gamma$ \\\midrule

\mbox{Scale-Free }500 &        500 &        499 &         14 &        4,3 &       3,03 &       1,27 &       2,24 \\

\mbox{Scale-Free }2500 &       2500 &       2499 &         21 &       4,25 &       3,54 &       0,71 &       2,83 \\

\mbox{Scale-Free }5000 &       5000 &       4999 &         22 &       4,77 &       3,59 &       1,18 &       2,46 \\

\mbox{Ring regular }$p=0$ &       5000 &     250000 &         50 &       1,02 &       1,02 &          0 &          $-$ \\

\mbox{Small World }$p=10^{-5}$ &       5000 &     250009 &         24 &       1,96 &       1,85 &       0,11 &          $-$ \\

\mbox{Small World }$p=2\cdot 10^{-5}$ &       5000 &     250020 &         22 &       2,17 &       2,05 &       0,12 &          $-$ \\

\mbox{Small World }$p=3\cdot 10^{-5}$ &       5000 &     250032 &         14 &       2,36 &       2,38 &       0,02 &          $-$ \\

\mbox{Random }$p=7\cdot 10^{-4}$ &       1043 &     1393 &         40 &       3,28 &       2,92 &       0,36 &                  $-$ \\

\mbox{Random }$p=4\cdot 10^{-4}$ &       2635 &     3144 &         40 &       4,21 &       3,64 &       0,57 &                  $-$ \\

\mbox{Random }$p=2\cdot 10^{-4}$ &       7924 &     9467 &         30 &       6,78 &       5,40 &       1,38 &                  $-$ \\

\bottomrule
\end{tabular}

    \caption{Parameters of the synthetic networks.}
    \label{tabla_sinteticas}
\end{table}

\section{Application in real networks}

In this subsection, the suitability of the Fixed-mass algorithm is explored for real networks.

\subsection{Power Grid}

The electrical Power Grid of the Western United States is considered. The power grid network \cite{Watts98} (available at \url{http://www-personal.umich.edu/~mejn/netdata/power.zip}) represents the topology of the Western States Power Grid of the United States. According to \cite{Phadke88}, this network is relevant to the efficiency and robustness of power networks. Its 4941 nodes represent generators, transformers and substations and its 13188 edges represent high-voltage transmission lines among them. Since the high-voltage transmission measure is not taken into account,this network is unweighted, i.e., the edges exclusively represent connections among adjacent the nodes. Hence, the weights considered for every edge might be the same value (e.g. weights equal to $1$).

\begin{figure}[H]
 	\begin{center}
 		\includegraphics[width=\textwidth]{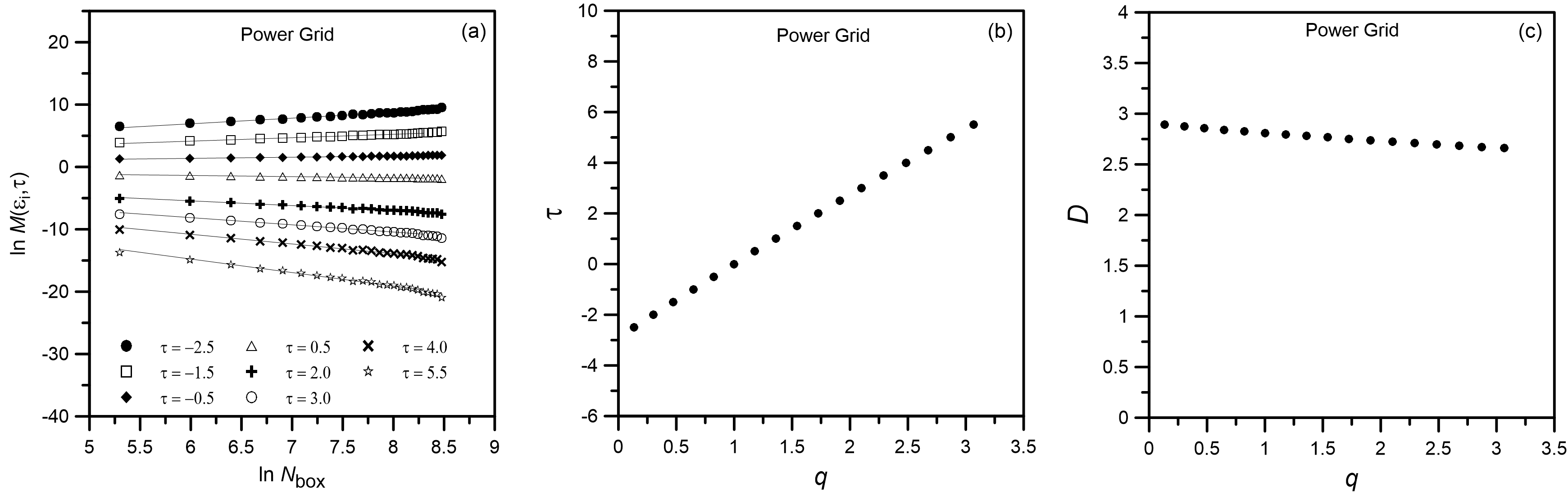}
         \caption{\textbf{Power Grid:} (a) Partition functions $\ln(M\bigl(\varepsilon_i,\tau)\bigr)$ versus $\ln(N_{box})$ for every value of $\tau$. (b) Exponent mass function. (c) Generalised dimension function.}
         \label{Fig_07}
     \end{center}
 \end{figure}

The proposed FMA is employed in this real network. The partition functions in Figure \ref{Fig_07}(a) shows a linear scaling relationship in a very a wide range, that covers practically all the  masses, from $N_{box} = 200$ to $N_{box} = 4000$. In this region, linear fits have been conducted with a very reliable goodness-fit $R^2 > 0.99$ and errors of estimation of the slope $(1-q) < 0.04$. From these linear fits, the mass exponent function is computed and depicted in Figure \ref{Fig_07}(b). On the other hand, the dimension function, $D_q$ shows a quasi-independence relation versus $q$, which evinces the tend to monofractallity of this complex network, which reaches a fractal dimension of 2.86 (\ref{Fig_07}(c)). Numerical results are summarized in Table \ref{tabla_reales}.

\subsection{C.Elegans \& S.Cerevisiae}

After applying the FMA to a real monofractal complex network, we are interested on testing the proposed algorithm to characterize multifractal complex networks. Two protein–protein interaction (PPI) data were computed in order to generate two unweighted complex networks, which are known to exhibit a multifractal behaviour (\cite{Wang2012}). We analyze the C. Elegans PPI network, obtained from \url{http://deim.urv.cat/~alexandre.arenas/data/welcome.htm} \cite{Duch05} and S. Cerevisiae (baker’s yeast) obtained from \url{http://vlado.fmf.uni-lj.si/pub/networks/data/bio/Yeast/Yeast.htm} \cite{Bu03}.

\begin{figure}[H]
 	\begin{center}
 		\includegraphics[width=\textwidth]{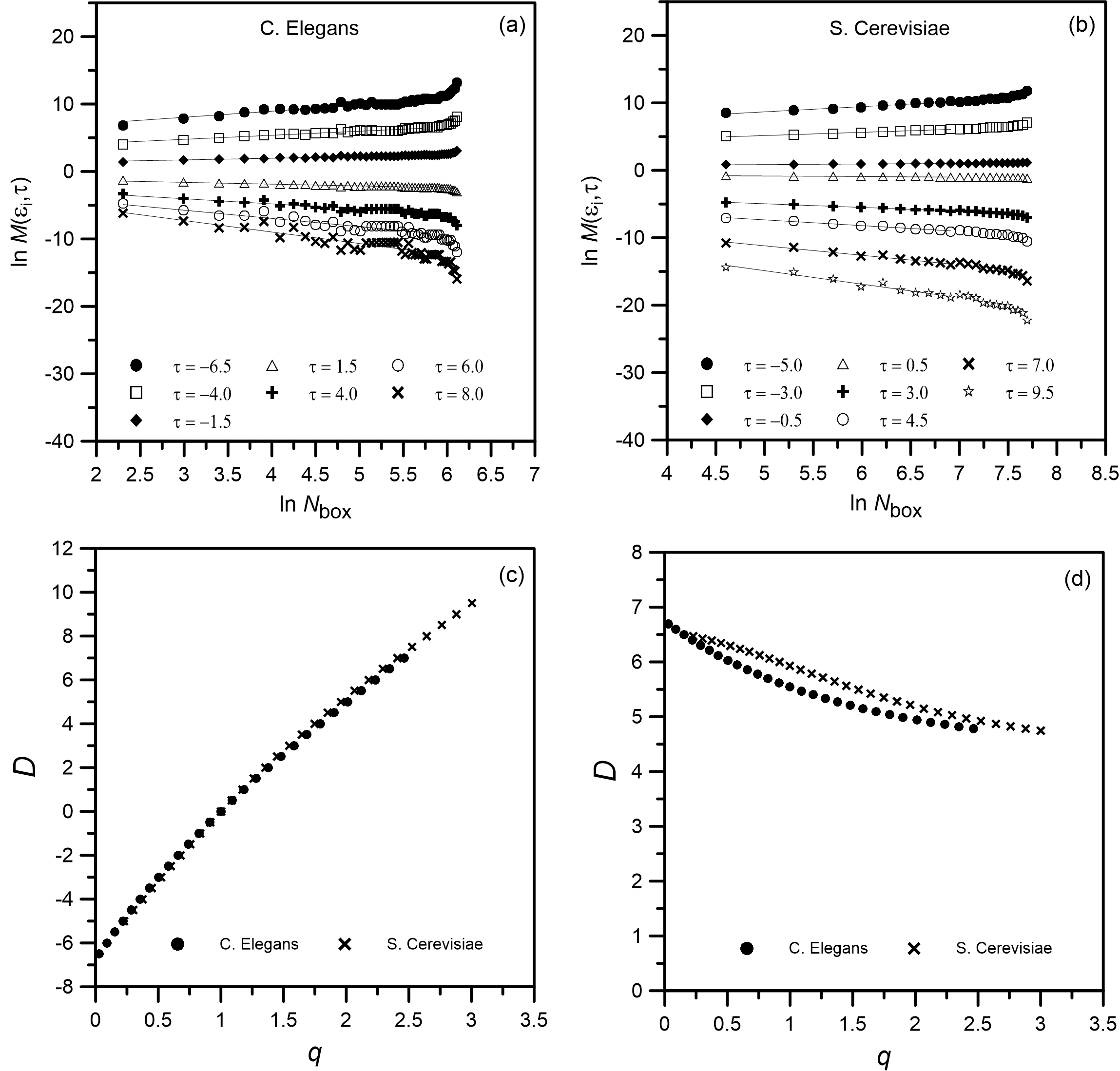}
         \caption{\textbf{C. Elegans and S. Cerevisiae networks:} Partition functions $\ln(M\bigl(\varepsilon_i,\tau)\bigr)$ versus $\ln(N_{box})$ for every value of $\tau$. (a) C. Elegans. (b) S. Cerevisiae. (c) Exponent mass functions. (d) Generalised dimension functions.}
         \label{Fig_08}
     \end{center}
 \end{figure}

According to Figure \ref{Fig_08}(a-b), partition functions of these PPI networks follow a similar behaviour for higher masses than scale-free networks. This evinces that the proposed FMA is capable to reproduce the effect detected in scale free network analysis, in which a few increment of mass in boxes is related to high increments on the diameter of the boxes. For C. Elegans, the linear fits are established from $N_{box}= 10$ to $N_{box}= 30$. In this region, linear fits have been conducted with $R^2 > 0.80$ and errors of estimation of the slope $(1-q)<0.21$. On the other hand, for S. Cerevisae the linear fits are fixed from $N_{box} = 100$ to $N_{box} = 1 000$, with $R^2 > 0.95$ and error $> 0.15$. From these linear fits, the mass exponent functions are computed and depicted in Figure \ref{Fig_08}(c). In \ref{Fig_08}(d) the dimension functions of both PPI networks are depicted, in which the multifractal behaviour is evinced by means of the typical decay of $D(q)$ versus $q$. Table \ref{tabla_reales} summarizes these multifractal results.

\begin{table}[H]
    \centering
    \begin{tabular}{lccccccc}
\toprule
 Network     & $N_{network}$ & $N_{edges}$ &  Diameter &  $D(q=0)$ & $D(q=3)$ & $\Delta(D_q)$ & $q_{range}$ \\\midrule
Powergrid    &        4941 &        6594 &         46 &        2.86 &       2.63 &       0.23 &      $[0,3]$ \\
C. Elegans   &       452 &       2010 &        8 &       6.69 &       4.71 &       1.98 &       $[0,2.6]$ \\
S. Cerevisae &      2224 &       6609 &         11 &       6.47 &       4.74 &       1.73 &       $[0,3]$ \\
\bottomrule
\end{tabular}

    \caption{Parameters of the real networks.}
    \label{tabla_reales}
\end{table}

\section{Conclusions}

In the last years, several algorithms have been developed for the multifractal analysis of different kinds of complex networks. However, researches carried out on this field have exclusively been restricted to the application of multifractal algorithms from a Fixed-size approach (FSA). In this work, a proposal for a Fixed-mass algorithm (FMA) is presented and evaluated in different complex networks. Thus, the main contribution of this work is to first explore the applicability of a FMA for studying fractal and multifractal characteristics of unweighted complex networks.

First, the FMA is tested in deterministic networks. Results evince that the proposed FMA is very precise and exhibit very low estimation errors. Second, the FMA is applied to different synthetic networks (scale-free, small-world and random), for which multifractal results are in agreement with the ones previously obtained by applying FSA. Finally, the FMA is then carried out in real complex networks with satisfactory results. FMA is capable to reproduce the monofractal behaviour of the Power Grid network and the multifractallity of the PPI networks with aceptable goodness-fit, mainly from the linearity and smoothness of the partition functions that are generated. The Fixed-mass approach has the advantage of: i) knowing a priori the minimum configuration of boxes (ideal) and the possibility to compare them with the ones provided in the FMA, and ii) obtaining relevant information about some features of the complex networks from the shape of the partition functions. As a disadvantage, the FMA code for completely cover the network with the minimum number of non-overlapped boxes is more computationally complex, which implies more runtime.

Future works might generalize the FMA for the study of weighted complex networks. It could be also interesting to explore other procedures for the network covering, which is the main task in the analysis of complex networks. Finally, the literature about this topic has grown very fast in last years, and thus, a review work about the applicability, advantages and disadvantages of these methods could be very suitable.

\textbf{Acknowledgments.}

The authors would like to express their deepest gratitude towards the reviewers for their valuable suggestions and comments that helped improve the manuscript considerably.

This work was funded by the Spanish Ministry of Science, Innovation and Universities and ERDF under contracts RTI2018-100754-B-I00 (iSUN project) and PGC-101514-B-100; and supported by the promotion program of research and knowledge transfer of the University of C\'{a}diz (Project PR2017-082).


\begin{thebibliography}{52}
\providecommand{\natexlab}[1]{#1}
\providecommand{\url}[1]{\texttt{#1}}
\expandafter\ifx\csname urlstyle\endcsname\relax
  \providecommand{\doi}[1]{doi: #1}\else
  \providecommand{\doi}{doi: \begingroup \urlstyle{rm}\Url}\fi

\bibitem[Albert and Barab\'asi(2002)]{Barabasi02}
R.~Albert and A.L. Barab\'asi.
\newblock Statistical mechanics of complex networks.
\newblock \emph{Rev. Mod. Phys.}, 74:\penalty0 47--97, 2002.

\bibitem[Song et~al.(2005)Song, Havlin, and Makse]{Song2005}
C.~Song, S.~Havlin, and H.A. Makse.
\newblock Self-similarity of complex networks.
\newblock \emph{Nature}, 433:\penalty0 392–--395, 2005.
\newblock \doi{10.1038/nature03248}.

\bibitem[Gallos et~al.(2012)Gallos, Makse, and Sigman]{Gallos2012}
L.~K. Gallos, H.~A. Makse, and M.~Sigman.
\newblock A small world of weak ties provides optimal global integration of
  self-similar modules in functional brain networks.
\newblock \emph{Proceedings of the National Academy of Sciences of the United
  States of America}, 109\penalty0 (8):\penalty0 2825--2830, 2012.
\newblock \doi{10.1073/pnas.1106612109}.

\bibitem[Watanabe et~al.(2015)Watanabe, Mizutaka, and Yakubo]{Watanabe2015}
A.~Watanabe, S.~Mizutaka, and K.~Yakubo.
\newblock Fractal and small-world networks formed by self-organized critical
  dynamics.
\newblock \emph{Journal of the Physical Society of Japan}, 84\penalty0
  (11):\penalty0 114003, 2015.
\newblock \doi{10.7566/JPSJ.84.114003}.

\bibitem[Koorehdavoudi and Bogdan()]{Bogdan2016}
H.~Koorehdavoudi and P.~Bogdan.
\newblock A statistical physics characterization of the complex systems
  dynamics: Quantifying complexity from spatio-temporal interactions.
\newblock \emph{Scientific Reports}, 6\penalty0 (27602).

\bibitem[Balaban et~al.()Balaban, Lim, Gupta, Boedicker, and
  Bogdan]{Bogdan2018}
V.~Balaban, S.~Lim, G.~Gupta, J.~Boedicker, and P.~Bogdan.
\newblock Quantifying emergence and self-organisation of enterobacter cloacae
  microbial communities.
\newblock \emph{Scientific Reports}, 8\penalty0 (12416).

\bibitem[Kawasaki and Yakubo(2010)]{Kawasaki10}
F.~Kawasaki and K.~Yakubo.
\newblock Reciprocal relation between the fractal and the small-world
  properties of complex networks.
\newblock \emph{Phys. Rev. E}, 82:\penalty0 036113, 2010.

\bibitem[Song et~al.(2006)Song, Havlin, and Makse]{Song2006}
C.~Song, S.~Havlin, and H.A. Makse.
\newblock Origins of fractality in the growth of complex networks.
\newblock \emph{Nature Physics}, 2:\penalty0 275–--281, 2006.
\newblock \doi{10.1038/nphys266}.

\bibitem[Liu et~al.(2017)Liu, Wang, Yu, and Xie]{Liu17}
J.L. Liu, J.~Wang, Z.G. Yu, and X.H. Xie.
\newblock Fractal and multifractal analyses of bipartite networks.
\newblock \emph{Scientific Reports}, 7\penalty0 (45588), 2017.
\newblock \doi{10.1038/srep45588}.

\bibitem[Wen and Jiang(2018)]{Wen2018}
T.~Wen and W.~Jiang.
\newblock {An information dimension of weighted complex networks}.
\newblock \emph{Physica A: Statistical Mechanics and its Applications},
  501\penalty0 (C):\penalty0 388--399, 2018.

\bibitem[Kim et~al.(2007{\natexlab{a}})Kim, Goh, Kahng, and Kim]{Kim2007}
J.~S. Kim, K.-I. Goh, B.~Kahng, and D.~Kim.
\newblock A box-covering algorithm for fractal scaling in scale-free networks.
\newblock \emph{Chaos: An Interdisciplinary Journal of Nonlinear Science},
  17\penalty0 (2):\penalty0 026116, 2007{\natexlab{a}}.
\newblock \doi{10.1063/1.2737827}.

\bibitem[Liu et~al.(2015)Liu, Yu, and Anh]{Liu2015}
J.L. Liu, Z.G. Yu, and V.~Anh.
\newblock Determination of multifractal dimensions of complex networks by means
  of the sandbox algorithm.
\newblock \emph{Chaos: An Interdisciplinary Journal of Nonlinear Science},
  25\penalty0 (2):\penalty0 023103, 2015.
\newblock \doi{10.1063/1.4907557}.

\bibitem[Feder(2013)]{feder88}
J.~Feder.
\newblock \emph{Fractals}.
\newblock Physics of Solids and Liquids. Springer US, 2013.
\newblock ISBN 9781489921246.

\bibitem[Furuya and Yakubo(2011)]{Furuya11}
S.~Furuya and K.~Yakubo.
\newblock Multifractality of complex networks.
\newblock \emph{Phys. Rev. E}, 84:\penalty0 036118, 2011.

\bibitem[Song et~al.(2007)Song, Gallos, Havlin, and Makse]{Song2007}
C.~Song, L.~K. Gallos, S.~Havlin, and H.~A. Makse.
\newblock How to calculate the fractal dimension of a complex network: the box
  covering algorithm.
\newblock \emph{Journal of Statistical Mechanics: Theory and Experiment}, 2007,
  2007.

\bibitem[Kim et~al.(2007{\natexlab{b}})Kim, Goh, Salvi, Oh, Kahng, and
  Kim]{Kim2007b}
J.~S. Kim, K.-I. Goh, G.~Salvi, E.~Oh, B.~Kahng, and D.~Kim.
\newblock Fractality in complex networks: Critical and supercritical skeletons.
\newblock \emph{Phys. Rev. E}, 75:\penalty0 016110, 2007{\natexlab{b}}.

\bibitem[Rozenfeld et~al.(2007)Rozenfeld, Havlin, and ben
  Avraham]{Rozenfeld2007}
H.~D. Rozenfeld, S.~Havlin, and D.~ben Avraham.
\newblock Fractal and transfractal recursive scale-free nets.
\newblock \emph{New Journal of Physics}, 2007.

\bibitem[Zhou et~al.(2007)Zhou, Jiang, and Sornette]{Zhou2007}
W.-X. Zhou, Z.-Q. Jiang, and D.~Sornette.
\newblock Exploring self-similarity of complex cellular networks: The
  edge-covering method with simulated annealing and log-periodic sampling.
\newblock \emph{Physica A: Statistical Mechanics and its Applications},
  375\penalty0 (2):\penalty0 741 -- 752, 2007.
\newblock ISSN 0378-4371.
\newblock \doi{10.1016/j.physa.2006.10.025}.

\bibitem[Gao et~al.(2008)Gao, Hu, and Di]{Gao2008}
L.~Gao, Y.~Hu, and Z.~Di.
\newblock Accuracy of the ball-covering approach for fractal dimensions of
  complex networks and a rank-driven algorithm.
\newblock \emph{Phys. Rev. E}, 78:\penalty0 046109, 2008.

\bibitem[Sun and Zhao(2014)]{Sun&Zhao2014}
Y.~Sun and Y.~Zhao.
\newblock Overlapping-box-covering method for the fractal dimension of complex
  networks.
\newblock \emph{Phys. Rev. E}, 89:\penalty0 042809, 2014.

\bibitem[Wang et~al.(2012)Wang, Yu, and Anh]{Wang2012}
D.L. Wang, Z.G. Yu, and V.~Anh.
\newblock Multifractal analysis of complex networks.
\newblock \emph{Chinese Physics B}, 21\penalty0 (8):\penalty0 080504, 2012.

\bibitem[Li et~al.(2014)Li, Yu, and Zhou]{Li14}
B.G. Li, Z.G. Yu, and Y.~Zhou.
\newblock Fractal and multifractal properties of a family of fractal networks.
\newblock \emph{Journal of Statistical Mechanics: Theory and Experiment},
  2014\penalty0 (2):\penalty0 P02020, 2014.

\bibitem[T\'el et~al.(1989)T\'el, F{\"u}l{\"o}p, and Vicsek]{Tel1989}
T.~T\'el, A.~F{\"u}l{\"o}p, and T.~Vicsek.
\newblock Determination of fractal dimensions for geometrical multifractals.
\newblock \emph{Physica A: Statistical Mechanics and its Applications},
  159\penalty0 (2):\penalty0 155 -- 166, 1989.
\newblock \doi{10.1016/0378-4371(89)90563-3}.

\bibitem[Vicsek(1990)]{VICSEK1990}
T.~Vicsek.
\newblock Mass multifractals.
\newblock \emph{Physica A: Statistical Mechanics and its Applications},
  168\penalty0 (1):\penalty0 490 -- 497, 1990.

\bibitem[Vicsek et~al.(1990)Vicsek, Family, and Meakin]{VicsekFamily1990}
T.~Vicsek, F.~Family, and P.~Meakin.
\newblock Multifractal geometry of diffusion-limited aggregates.
\newblock 1990.

\bibitem[Wei et~al.(2013)Wei, Liu, Zhang, Hu, Deng, and Mahadevan]{Wei2013}
D.J. Wei, Q.~Liu, H.X. Zhang, Y.~Hu, Y.~Deng, and S.~Mahadevan.
\newblock Box-covering algorithm for fractal dimension of weighted networks.
\newblock \emph{Scientific Reports}, 3\penalty0 (3049), 2013.
\newblock \doi{10.1038/srep03049}.

\bibitem[Song et~al.(2015)Song, Liu, Yu, and Li]{Song2015}
Y.Q. Song, J.L. Liu, Z.G. Yu, and B.G. Li.
\newblock Multifractal analysis of weighted networks by a modified sandbox
  algorithm.
\newblock \emph{Scientific Reports}, 5\penalty0 (17628), 2015.

\bibitem[Wei et~al.(2016)Wei, Chen, and Deng]{Wei2016}
D.~Wei, X.~Chen, and Y.~Deng.
\newblock Multifractality of weighted complex networks.
\newblock \emph{Chinese Journal of Physics}, 54\penalty0 (3):\penalty0 416 --
  423, 2016.

\bibitem[Xue and Bogdan(2017)]{Xue17}
Y.~Xue and P.~Bogdan.
\newblock Reliable multi-fractal characterization of weighted complex networks:
  Algorithms and implications.
\newblock \emph{Scientific Reports}, 7:\penalty0 7487, 2017.
\newblock \doi{10.1038/s41598-017-07209-5}.

\bibitem[Theiler()]{Theiler90}
J.~Theiler.
\newblock Estimating fractal dimension.
\newblock \emph{J. Opt. Soc. Am. A}, 7\penalty0 (6):\penalty0 1055--1073.

\bibitem[Termonia and Alexandrowicz(1983)]{Termonia83}
Y.~Termonia and Z.~Alexandrowicz.
\newblock Fractal dimension of strange attractors from radius versus size of
  arbitrary clusters.
\newblock \emph{Phys. Rev. Lett.}, 51:\penalty0 1265--1268, 1983.

\bibitem[Badii and Politi(1984)]{Badii1984}
R.~Badii and A.~Politi.
\newblock Hausdorff dimension and uniformity factor of strange attractors.
\newblock \emph{Phys. Rev. Lett.}, 52:\penalty0 1661--1664, 1984.

\bibitem[Badii and Politi(1985)]{Badii1985}
R.~Badii and A.~Politi.
\newblock Statistical description of chaotic attractors: The dimension
  function.
\newblock \emph{Journal of Statistical Physics}, 1985.

\bibitem[Badii and Broggi(1988)]{BadiiBroggi88}
R.~Badii and G.~Broggi.
\newblock Measurement of the dimension spectrum f($\alpha$): Fixed-mass
  approach.
\newblock \emph{Physics Letters A}, 131\penalty0 (6):\penalty0 339--343, 1988.
\newblock \doi{10.1016/0375-9601(88)90784-0}.

\bibitem[Grassberger et~al.(1988)Grassberger, Badii, and
  Politi]{Grassberger1988}
P.~Grassberger, R.~Badii, and A.~Politi.
\newblock Scaling laws for invariant measures on hyperbolic and nonhyperbolic
  atractors.
\newblock \emph{Journal of Statistical Physics}, 1988.

\bibitem[Hirabayashi et~al.(1992)Hirabayashi, Ito, and Yoshii]{Hirabayashi1992}
T.~Hirabayashi, K.~Ito, and T.~Yoshii.
\newblock Multifractal analysis of earthquakes.
\newblock \emph{Pure and applied geophysics}, 1992.

\bibitem[Mach et~al.(1995)Mach, Mas, and Sagu\'{e}s]{Mach95}
J.~Mach, F.~Mas, and F.~Sagu\'{e}s.
\newblock Two representations in multifractal analysis.
\newblock \emph{Journal of Physics A: Mathematical and General}, 28\penalty0
  (19):\penalty0 5607, 1995.

\bibitem[Pastor-Satorras(1997)]{Pastor97}
R.~Pastor-Satorras.
\newblock Multifractal properties of power-law time sequences: application to
  rice piles.
\newblock \emph{Physical review E, statistical physics, plasmas, fluids, and
  related interdisciplinary topics}, 1997.

\bibitem[Bartolo et~al.(2000)Bartolo, Gabriele, and Gaudio]{DeBartolo00}
S.G.~De Bartolo, S.~Gabriele, and R.~Gaudio.
\newblock Multifractal behaviour of river networks.
\newblock \emph{Hydrology and Earth System Sciences}, 4\penalty0 (1):\penalty0
  105--112, 2000.
\newblock \doi{10.5194/hess-4-105-2000}.

\bibitem[Enescu et~al.(2005)Enescu, Ito, Radulian, E.Popescu, and
  Bazacliu]{Enescu2005}
B.~Enescu, K.~Ito, M.~Radulian, E.Popescu, and O.~Bazacliu.
\newblock Multifractal and chaotic analysis of vrancea (romania)
  intermediate-depth earthquakes: Investigation of the temporal distribution of
  events.
\newblock \emph{Pure and applied geophysics}, 2005.

\bibitem[Behnia et~al.(2012)Behnia, Akhshani, Panahi, Mobaraki, and
  Chaderian]{Behnia12}
S.~Behnia, A.~Akhshani, M.~Panahi, A.~Mobaraki, and M.~Chaderian.
\newblock Multifractal properties of denaturation process based on
  peyrard-bishop model.
\newblock \emph{Physics Letters A}, 376\penalty0 (37):\penalty0 2538–--2547,
  2012.
\newblock \doi{10.1016/j.physleta.2012.05.062}.

\bibitem[Falconer(1994)]{Falconer94}
K.J. Falconer.
\newblock The multifractal spectrum of statistically self-similar measures.
\newblock \emph{Journal of Theoretical Probability}, 1994.

\bibitem[Grassberger(1983)]{Grassberger83}
P.~Grassberger.
\newblock Generalized dimensions of strange attractors.
\newblock \emph{Physics Letters A}, 97\penalty0 (6):\penalty0 227 -- 230, 1983.
\newblock \doi{10.1016/0375-9601(83)90753-3}.

\bibitem[Bartolo et~al.(2006)Bartolo, Primavera, Gaudio, D'Ippolito, and
  Veltri]{FmaDeBartolo06}
S.G.~De Bartolo, L.~Primavera, R.~Gaudio, A.~D'Ippolito, and M.~Veltri.
\newblock Fixed-mass multifractal analysis of river networks and braided
  channels.
\newblock \emph{Physical Review E}, 74:\penalty0 026101, 2006.
\newblock \doi{10.1103/PhysRevE.74.026101}.

\bibitem[Meneveau and Sreenivasan(1991)]{meneveau_sreenivasan_1991}
C.~Meneveau and K.~R. Sreenivasan.
\newblock The multifractal nature of turbulent energy dissipation.
\newblock \emph{Journal of Fluid Mechanics}, 224:\penalty0 429–484, 1991.
\newblock \doi{10.1017/S0022112091001830}.

\bibitem[Barab{\'a}si and Albert(1999)]{Barabasi99}
A.L. Barab{\'a}si and R.~Albert.
\newblock Emergence of scaling in random networks.
\newblock \emph{Science}, 286\penalty0 (5439):\penalty0 509--512, 1999.
\newblock \doi{10.1126/science.286.5439.509}.

\bibitem[Watts and Strogatz(1998)]{Watts98}
D.J. Watts and S.H. Strogatz.
\newblock Collective dynamics of 'small-world' networks.
\newblock \emph{Nature}, 393:\penalty0 440--442, 1998.
\newblock \doi{doi:10.1038/30918}.

\bibitem[Newman and Watts(1999)]{Newman99}
M.E.J. Newman and D.J. Watts.
\newblock Renormalization group analysis of the small-world network model.
\newblock \emph{Physics Letters A}, 263\penalty0 (4):\penalty0 341 -- 346,
  1999.
\newblock ISSN 0375-9601.
\newblock \doi{10.1016/S0375-9601(99)00757-4}.

\bibitem[Erdős and R{\'e}nyi(1960)]{Erdos60}
P.~Erdős and A.~R{\'e}nyi.
\newblock On the evolution of random graphs.
\newblock pages 17--61, 1960.

\bibitem[Phadke and Thorp(1988)]{Phadke88}
A.G. Phadke and J.S. Thorp.
\newblock \emph{Computer Relaying for Power Systems, Second Edition}.
\newblock Wiley, New York, 1988.

\bibitem[Duch and Arenas(2005)]{Duch05}
J.~Duch and A.~Arenas.
\newblock Community detection in complex networks using extremal optimization.
\newblock \emph{Phys. Rev. E}, 72:\penalty0 027104, 2005.

\bibitem[Bu et~al.(2003)Bu, Zhao, Cai, Xue, Zhu, Lu, Zhang, Sun, Ling, Zhang,
  Li, and Chen]{Bu03}
D.~Bu, Y.~Zhao, L.~Cai, H.~Xue, X.~Zhu, H.~Lu, J.~Zhang, S.~Sun, L.~Ling,
  N.~Zhang, G.~Li, and R.~Chen.
\newblock Topological structure analysis of the protein–protein interaction
  network in budding yeast.
\newblock \emph{Nucleic Acids Research}, 31\penalty0 (9):\penalty0 2443--2450,
  2003.

\end{thebibliography}

\appendix

\section{Fixed Mass algorithm for unweighted networks}

In this section we show the FMA MATLAB code.

\begin{lstlisting}

function [logmasses, logM] = FM_BC_unw(D,set_tau,set_masses,T)
%%%%%%%%%%%%%%%%%%%%%%%%%%%%%%%%%%%%%%%%%%%%%%%%%%%%%%%%%%%%%%%%%
%%%
%%% Fixed Mass algorithm for unweighted networks
%%%
%%%---------------------------------------
%%%
%%% INPUT: D          - Shortest path matrix.
%%%        set_tau    - Array with the values of tau.
%%%        set_masses - Array with the values of masses,
%%%                     between 2 and N_network.
%%%        T          - Number of realizations.
%%%
%%% OUTPUT: logmasses - Array with the logarithm of the masses.
%%%         logM      - Matrix with M(eps_i,tau)
%%%
%%% -------------------------------------
%%%
%%%%%%%%%%%%%%%%%%%%%%%%%%%%%%%%%%%%%%%%%%%%%%%%%%%%%%%%%%%%%%%%%

N_network = length(D);

for t=1:T  %%% Step (2): Node rearrangement
    TT(t,:)=randperm(N_network,N_network);
end

covered = [];

B = zeros(length(set_masses),T); %% Create the matrix B(N_box,t)

%%%% Steps (4)-(6): Create the boxes

for N_box_index = 1:length(set_masses)

    N_box = set_masses(N_box_index);
    n_boxes = 0;

    for t=1:T   %% Run through every realization
        for counter=1:N_network
            center = TT(t,counter);
            neighbours_box = [];
            flag = 0; %% Inicialize a flag. If flag is zero, the box is discarded.
            distance = 1; %% Inicialize a distance.

            %% Search for N_box - 1 neighbours

            while length(neighbours_box) <= (set_masses(N_box_index)-1) & (distance<=set_masses(N_box_index)-1)
                neighbours = find(D(center,:) == distance);
                not_covered_neighbours = neighbours(ismember(neighbours,covered)~=1);
                covered_neighbours = neighbours(ismember(neighbours,covered)~=0);

                %%% A rearrangement is introduced in order to choose
                %%% nodes at the last distance randomly
                not_covered_neighbours = not_covered_neighbours(randperm(length(not_covered_neighbours),
                                         length(not_covered_neighbours)));
                covered_neighbours = covered_neighbours(randperm(length(covered_neighbours),
                                     length(covered_neighbours)));

                if not_covered_neighbours ~= 0 %% If there are not covered neighbours, activate a flag. Because a new box will be created.
                    flag = 1;
                end

                neighbours_box = [neighbours_box not_covered_neighbours covered_neighbours];
                %%% Notice that the algorithm prioritizes the new not covered nodes over the covered ones.
                distance = distance + 1; %%% Increase the distance.
            end
               %%% If flag = 0, the box is discarded
            if flag == 1   %%% If flag = 1, the box is created
                box = [center neighbours_box(1:set_masses(N_box_index) - 1)];
                n_boxes = n_boxes +1; %%% Count the number of boxes.
                covered = unique([covered box]); %%% Save the covered nodes
                n_covered = length(covered); %%% Count the covered nodes
                box_diameter(N_box_index,n_boxes,t) = max(max(D(box,box))); %%% Retain the diameter of the box (subgraph)
            end

            if n_covered==N_network %%% Stop if all nodes are covered.
                break
            end
        end
        B(N_box_index,t) = n_boxes;
        %%%% Reset the parameters
        n_boxes = 0;
        covered=[];
    end
end

%%% Step (7): Retain the minimum number of boxes for each mass. This is an array.

B_min = min(B')';

%%%%% For each mass, save the realization with the minimum number of boxes.

for N_box_index =1:length(set_masses)
    [row,col] = find(B(N_box_index,:)==B_min(N_box_index));
    T_min(N_box_index)= col(1);
    clear row
    clear col
end
T_min = T_min';

%%%% MULTIFRACTAL ANALYSIS

for N_box_index=1:length(set_masses) %%% For each N_box, compute the diameter of each box
    diameter=box_diameter(N_box_index,:,T_min(N_box_index));
    diameter=diameter(diameter~=0);
    %%% Create the partition sum for every N_box
    for tau_index = 1:length(set_tau)
        tau = set_tau(tau_index);
        M(N_box_index,tau_index)=mean(diameter.^(-tau));
    end
    clear diameter
end

logmasses = log(set_masses)';
logM = log(M);
end

\end{lstlisting}

\end{document}